\documentclass[a4paper,11pt]{article}
\pdfoutput=1
\usepackage{jheppub}
\usepackage{amsthm}
\usepackage{amsmath}
\usepackage{float}
\usepackage{graphicx}
\usepackage{subcaption}
\usepackage{slashed}
\usepackage{amssymb}
\usepackage{xcolor}
\usepackage{soul}
\usepackage{comment}

\newcommand{\sac}{\, , \qquad}
\newcommand{\be}{\begin{eqnarray}}
\newcommand{\ee}{\end{eqnarray}}
\newcommand{\mpl}{M_\mathrm{p}}
\newcommand{\msp}{M_\mathrm{sp}}
\newcommand{\mspren}{M_\mathrm{sp, ren}}
\newcommand{\hcos}{H_{\rm cosmo}}
\newcommand{\mef}{M_\mathrm{sp}}

\newcommand{\beq}{\begin{equation}}
\newcommand{\eeq}{\end{equation}}
\newcommand{\hmin}{H_{\text{min}}}
\newcommand{\mm}{M}

\begin{document}

\title{Self-sustained, out-of-equilibrium inflation}

\author[1]{Jorge Casalderrey-Solana,}
\author[1]{Lucía Castells-Tiestos,}
\author[1]{Jéssica Gonçalves,}
\affiliation[1]{Departament de F\'\i sica Qu\`antica i Astrof\'\i sica \& Institut de Ci\`encies del Cosmos (ICC), 
Universitat de Barcelona, Mart\'{\i}  i Franqu\`es 1, 08028 Barcelona, Spain}

\author[1,2]{David~Mateos}
\affiliation[2]{Instituci\'o Catalana de Recerca i Estudis Avan\c cats (ICREA), Passeig Llu\'\i s Companys 23,  ES-08010, Barcelona, Spain.}

\emailAdd{jorge.casalderrey@ub.edu}
\emailAdd{lucia.castells@icc.ub.edu}
\emailAdd{jessica.goncalves@icc.ub.edu}
\emailAdd{dmateos@ub.edu}

\preprint{} 

\abstract{
We use holography to study dS-invariant states of non-conformal, strongly coupled quantum field theories in four-dimensional de Sitter space. We show that out-of-equilibrium 
effects can sustain the exponential inflation within the regime of validity of semiclassical gravity, $H \ll M \ll \msp$, with $H$  the Hubble parameter, $M$  the characteristic scale of the quantum field theory, $\msp = \mpl/N$  the species scale, $\mpl$  the Planck scale, and $N^2$  the number of matter fields. In the holographic description, the required fine-tuning scales only logarithmically with the ratio $\msp/H$. The resulting solutions exhibit apparent horizons whose increasing area indicates a continuous growth of the comoving entropy density. We  suggest that this inflationary regime can arise as the late-time limit of a dynamical evolution starting from an initial Friedmann–Lemaître–Robertson–Walker universe.
}

\maketitle

\section{Introduction}
Inflation~\cite{Starobinsky:1979ty,Starobinsky:1980te,Starobinsky:1982ee,Mukhanov:1981xt,Guth:1980zm,Linde:1981mu,Linde:1983gd,Albrecht:1982wi,Kofman:1985aw} is one of the most fascinating processes believed to have occurred in the early universe. Multiple cosmological observations support the conclusion that the universe underwent a brief period of accelerated expansion, during which physical distances increased by a factor of at least $10^{26}$. According to our current understanding, the spacetime geometry during this epoch was very close to that of de Sitter (dS) space. The end of inflation marked the onset of the hot Big Bang phase and the creation of matter in our universe.

The mechanisms underlying this remarkable phenomenon have been extensively studied over the past several decades (see~\cite{Lyth:1998xn,Guth:2013sya,Kallosh:2025ijd} for reviews). The most phenomenologically successful models typically invoke a fundamental scalar field—the inflaton—whose slow evolution along a nearly flat potential drives the accelerated expansion. 
In some realizations, this field is identified with the Higgs boson, supplemented by curvature-dependent couplings~\cite{Bezrukov:2007ep}, while other frameworks involve large-field~\cite{Linde:1983gd}, multi-field, or warm inflation~\cite{Berera:1995ie} scenarios. 
In certain cases, these scalar degrees of freedom can be viewed as effective descriptions of collective matter dynamics via the dynamics of scalar order parameters governing phase transitions, as in false-vacuum inflation. 
Although originally introduced to explain a later, non-primordial inflationary epoch, thermal inflation~\cite{Lyth:1995hj,Lyth:1995ka} also fits within this broader picture, with its onset understood as a response of matter to cosmological expansion. 

In this paper  we will show that exponential acceleration can arise simply from out-of-equilibrium properties of a non-conformal quantum field theory (QFT) in dS-invariant states, leading to self-sustained inflation: the expansion of the universe generates a nonzero stress-energy tensor that, in turn, sustains the expansion. This idea has a long history, dating back to the seminal work of Ref.~\cite{Starobinsky:1980te}, in which inflation is driven by the trace anomaly. The crucial difference between our work and  previous studies~\cite{Starobinsky:1980te,Vilenkin:1985md,Hawking:2000bb,Fabris:2000gz,Buchel:2002uw,Buchel:2002ya,Buchel:2003ax,Buchel:2004ku,Buchel:2006qp,Buchel:2014dva,Anguelova:2015dsv,Anguelova:2016usr,Buchel:2017pto,Casalderrey-Solana:2020vls}
is that,  in our case, this mechanism can be realized fully  within the regime of validity of semiclassical gravity:
\begin{equation}
    H\ll M \ll \msp \,.
    \label{semi}
\end{equation}
Moreover, the required fine-tuning of the model parameters grows only logarithmically with the ratio $\msp/H$.

In~\eqref{semi}, $H$ denotes the expansion rate and  $M$  the characteristic mass scale of the QFT. The latter encodes the breaking of conformal invariance, which is essential for our mechanism: in a conformal field theory (CFT) on dS space the stress tensor is completely fixed by the anomaly, leaving no room for the dynamics required here.
The  species scale, $\msp$, is related to the  Planck mass and to the number of matter fields $N^2$ through  
\begin{equation}
\label{species}
    \msp = \frac{\mpl}{N} \,.
\end{equation}
Although we believe our mechanism is general, in order to have calculational control we will be interested in the limit $N\to\infty$. This must be taken keeping $\msp$ fixed while sending $\mpl\to\infty$. This results in a well defined semiclassical gravitational theory in which quantum gravity effects become important at the species scale~\cite{Dvali:2007hz,Chesler:2020exl,Ghosh:2023gvc,Castellano:2022bvr}. The second inequality in~\eqref{semi} is simply the requirement that the QFT scale is sufficiently lower than the quantum gravity scale. 
At large $N$, the energy density of the QFT scales as $N^2 \mathcal{E}$, with $\mathcal{E}$ of order $N^0$.  
Assuming that semiclassical gravity holds up to characteristic energy densities in the QFT of order $\mathcal{E} \sim M^4$, the first inequality in~\eqref{semi} follows from the Friedmann equation 
\beq
\label{fri}
H^2 \sim \frac{\mathcal{E}}{\msp^2} \,.
\eeq
In summary, treating gravity classically is only justified if the hierarchy~\eqref{semi} is satisfied. 

Much of our current understanding of QFT in curved spacetime relies on analyses of free or weakly interacting fields in fixed gravitational backgrounds (see~\cite{Birrell:1982ix} and references therein). In recent years, however, the framework has been extended to the strongly coupled regime, primarily through holography~\cite{Marolf:2010tg,Penin:2021sry,Buchel:2019qcq,Buchel:2017lhu,Buchel:2017pto,Buchel:2016cbj,Casalderrey-Solana:2020vls,Ecker:2021cvz,Mashayekhi:2025jyg,Kastikainen:2025eys,Ghosh:2017big,Apostolopoulos:2008ru,Koyama:2001rf}. Here we  follow the latter approach and determine the dS-invariant states of a  one-parameter family of strongly-coupled, \mbox{large-$N$} QFTs in a fixed, four-dimensional dS spacetime. 
Their dual description, henceforth referred to as ``the bulk'', is given by five-dimensional gravity coupled to a scalar field in an asymptotically anti–de Sitter (AdS) spacetime. Different QFTs in dS correspond to different choices of the  scalar-field potential in  the bulk. 

\begin{figure}[t]
    \begin{center}
    \includegraphics[width=0.8\textwidth]{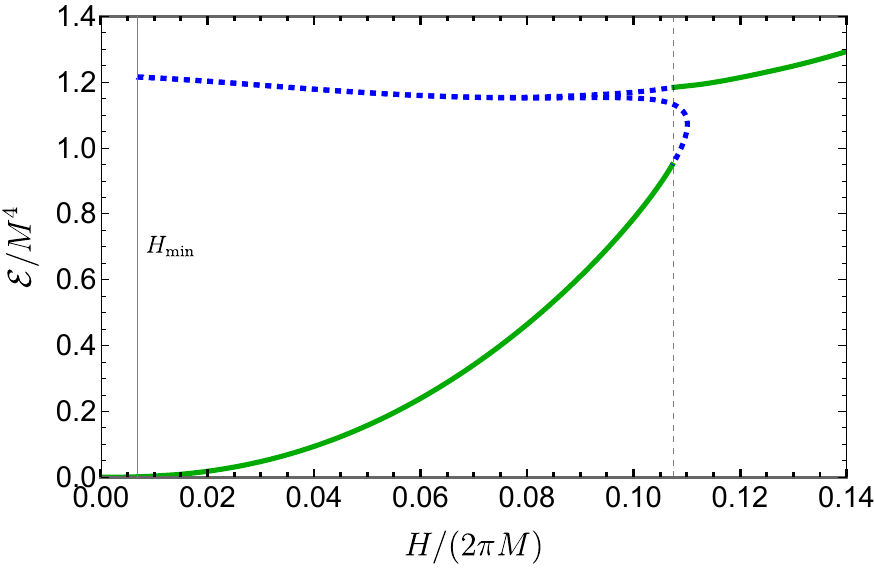}
    \end{center}
    \caption{\small\label{Edens059_D}
   Energy density of dS-invariant states of a QFT in our family (corresponding to $\phi_M=0.59$) as a function of the dS Hubble parameter. Multiple states coexist for the same value of $H$ down to a small value $\hmin \ll M$, indicated by a solid vertical line on the left, with a finite energy gap between them. States with the lowest Euclidean action are shown in solid green. The dashed vertical line on the right indicates the point at which these states jump from one branch to another. The blue branches lie very close to each other and intersect only at $H=\hmin$.} 
\end{figure}

Our main result is that, for appropriate choices of the potential, the energy density of the QFT takes the qualitative form shown in Fig.~\ref{Edens059_D}.
The important feature is that, in a certain region that extends down to a value $\hmin \ll M$ marked by the solid vertical line, three different vacua exist for each value of $H$: one lies on the lower green branch and the other two lie on the blue branches. The latter lie very close to each other and intersect only at $H=\hmin$.  
The green branch extends all the way to $H=0$. The blue  branches do not. Therefore, these branches are  invisible to experiments that study  small perturbations around flat space. 

Up to logarithmic corrections,  the energy density in the lower green branch at small $H$ scales as $O(H^2 M^2)$. As we will discuss in Section~\ref{sec:cosmo}, the stress tensor in this branch cannot self-consistently support 
the accelerated expansion in the semiclassical regime, for reasons analogous to those in the trace anomaly-driven inflationary scenario of~\cite{Starobinsky:1980te}.\footnote{Despite this, Starobinsky inflation has interesting  phenomenological applications \cite{Mukhanov:1981xt,Lyth:1998xn,German:2023euc}.}

In contrast, the energy density at small $H$ in the blue branches scales as $O(M^4)$. This leads to self-sustained inflation within the semiclassical regime~\eqref{semi} with
$H\gtrsim \hmin$,  provided $\hmin$ is sufficiently small compared to $M$, i.e.
\beq
\label{requi}
\frac{\hmin}{M} \lesssim \frac{M}{\msp} \,.
\eeq
This condition clearly requires fine-tuning. From the QFT perspective, this may appear highly unnatural due to the large ratio $\msp/M$. Nevertheless, we will show that, in the five-dimensional  description, the required fine-tuning scales only logarithmically with this ratio. Therefore, in the dual description, what might appear fine-tuned from the QFT perspective can in fact be satisfied rather naturally. As we will explain, the underlying reason is the same universal property of AdS that makes the Randall–Sundrum solution to the hierarchy problem~\cite{Randall:1999ee} possible: the AdS warp factor grows exponentially with the proper distance along the holographic direction.

The stress-energy tensor of quantum fields in dS space is inherently ambiguous due to scheme dependence.  However, the energy difference between the blue curves and the green curve is scheme-independent, and is therefore physically observable.
Moreover, these ambiguities arise from treating gravity as an external, non-dynamical field. Although our calculations are performed in a fixed dS background, imposing the Friedmann equation requires coupling the QFT to dynamical gravity. As we will review in Section~\ref{scheme}, in this case the scheme-dependent terms in the stress  tensor combine with the bare parameters in the gravitational Lagrangian to yield the renormalized, physical values of the cosmological constant and the species scale. Within this framework, the discussion surrounding Fig.~\ref{Edens059_D} becomes scheme-independent and depends only on one physical assumption: that the lower, green branch connects smoothly to flat space, namely, that flat space is a solution of the  QFT coupled to dynamical  gravity. Under this assumption, coupling the vacuum states on the blue branches to dynamical gravity results in self-sustained inflation within the semiclassical 
regime.

We strongly emphasize that we are not claiming to have a phenomenologically viable model of inflation. Our goal in this paper is more modest: to present a setup fully under theoretical control where accelerated expansion is driven entirely by out-of-equilibrium QFT effects.
While accelerated expansion is a necessary ingredient for inflation, it is not sufficient on its own. Assessing whether this scenario could function as an inflationary model would require, among other aspects, an analysis of cosmological perturbations, mechanisms for a graceful exit, and a reheating process. While we leave these investigations to future work, we can comment on a possible route into the inflationary regime described here.

Indeed, Refs.~\cite{Buchel:2017pto,Casalderrey-Solana:2020vls} demonstrated explicitly  that non-conformal, strongly coupled matter in dS space evolves toward a state exhibiting exact dS symmetry---the strongly coupled analog of the Bunch–Davies vacuum. As the system approaches this attractor, the effects of expansion become increasingly significant, eventually driving the matter out of equilibrium. Beyond this point, the stress tensor  can no longer treated as that of an adiabatically-evolving  fluid governed by an equation of state derived from the thermodynamic properties of the QFT in flat space. Although in the late-time dS-invariant state the pressure and energy density become related again, this relation is enforced by dS-symmetry rather than by local thermal equilibrium. Consequently, the evolution  should be accompanied by  entropy production. Remarkably, this is confirmed by the analysis in~\cite{Buchel:2017pto}: the holographic duals of these non-conformal dS–symmetric states feature  time-dependent apparent horizons with growing comoving area density, signaling continuous entropy generation at the attractor solution. 

The key difference between the present work and Refs.~\cite{Buchel:2017pto,Casalderrey-Solana:2020vls} is the choice of scalar potential in the bulk, which in those models does not give rise to the multiple branches displayed in Fig.~\ref{Edens059_D}. Nevertheless, as we discuss in Section \ref{sec:cosmo}, the results from those references suggest that the dS‑invariant states studied here could emerge from the dynamical evolution of a Friedmann–Lemaître–Robertson–Walker (FLRW) universe initially filled by a non-conformal, strongly coupled QFT fluid, with an intermediate phase of thermal inflation connecting the early FLRW stage and the late-time dS phase.

The family of holographic models that we study was originally introduced in~\cite{Bea:2018whf}. The physics of the dual QFTs in flat space has been extensively studied both in~\cite{Bea:2020ees,Ares:2021nap,Faedo:2024zib,Bea:2024xgv,Henriksson:2025vci} and out~\cite{Bea:2021zsu,Bea:2021ieq,Bea:2021zol,Bea:2022mfb,Escriva:2022yaf,Casalderrey-Solana:2023zlg} of equilibrium. For suitable parameter values, the thermal phase diagram in flat space develops  stable, metastable, and unstable branches at fixed temperature, with first-order transitions between them. As we elaborate below, this suggests the possibility of analogous curvature-driven phase transitions between the branches in Fig.~\ref{Edens059_D}, similar to those discussed in Ref.~\cite{Ghosh:2017big}, where they were termed “quantum phase transitions” because they are induced by curvature rather than thermal effects.

This paper is organized as follows. In Section~\ref{holographic-model}, we introduce the holographic setup and outline the methods used to characterize the strongly coupled states of the dual field theory. In Section~\ref{sec:fases}, we present a specific model that exhibits multiple dS–symmetric phases extending to expansion rates \(\hmin \ll \mm\), and we compare its curvature-driven and thermal phase structures. To further elucidate their similarities and differences, we perform a broader comparison across several representatives of the same family of models. In Section~\ref{sec:cosmo}, we explore the cosmological implications of these findings and argue that the small-curvature dS states naturally lead to a phase of out-from-equilibrium inflation. Finally, in Section~\ref{discussion}, we summarize our main results, discuss their limitations, and outline possible directions for future work.

\section{Holographic model}
\label{holographic-model}
\subsection{Action}
The family of strongly coupled QFTs that we study is defined via its holographic dual. On the gravity side, the system is described by an Einstein-scalar theory in five dimensions, with an action given by  
\begin{equation}
\label{eq:Sgen}
S=\frac{2}{8\pi G_5}\int d^{5}x\sqrt{-g}\left(\frac{1}{4}R[g]-\frac{1}{2}(\partial\phi)^{2}-V(\phi)\right)+\frac{1}{8\pi G_5}\int d^{4}x\sqrt{-\gamma}K+S_{ct}\;.
\end{equation}
Here, $G_5$ is the five-dimensional Newton constant, the second integral is the Gibbons-Hawking term, and $S_{ct}$ is the counterterm action required to ensure the finiteness of the total action. We will return to this term in the next section. 

\begin{figure}[t]
    \begin{center}
    \includegraphics[width=0.93\textwidth]{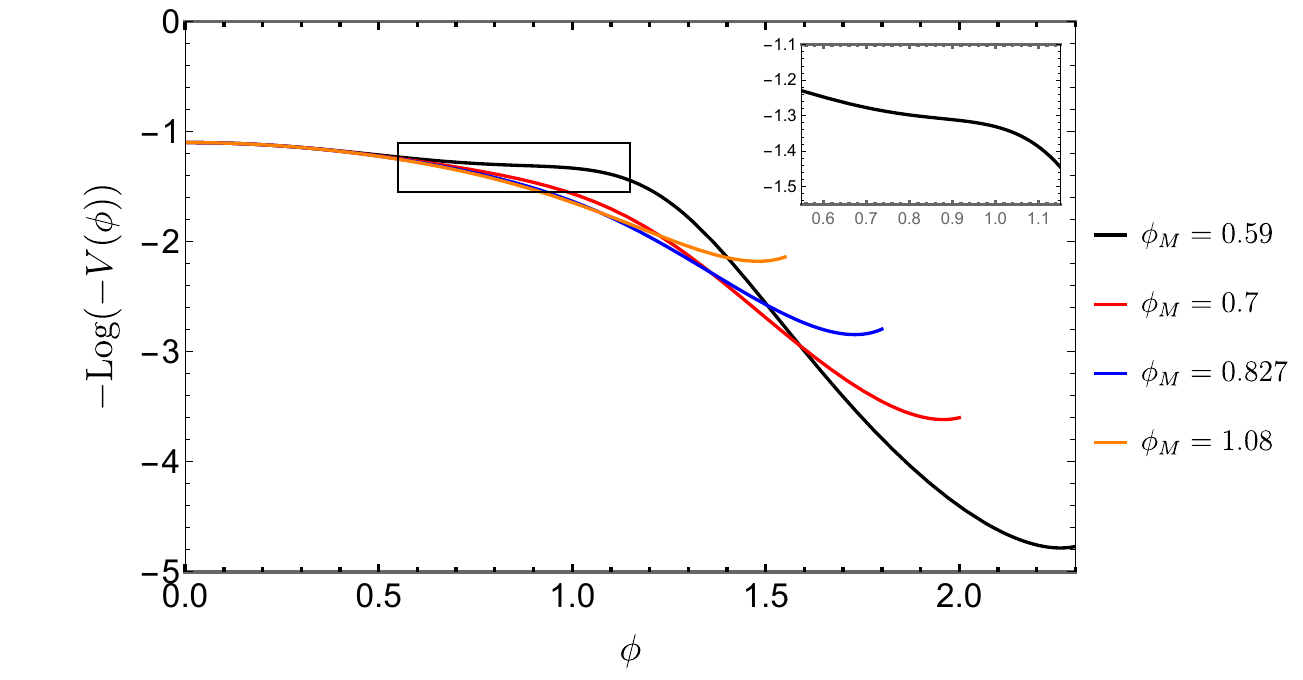}
    \end{center}
    \caption{\small\label{potpot}
    Bulk scalar potential for several values of $\phi_M$ that will be of interest below.} 
\end{figure}

The  action depends on the scalar potential $V(\phi)$. Since we are considering a bottom-up model, specifying $V(\phi)$ fully defines the theory. We focus on the one-parameter family of potentials introduced in \cite{Bea:2018whf} 
\be
\label{V}
V(\phi)=-\frac{4}{3}W(\phi)^{2}+\frac{W'(\phi)^{2}}{2}\;,
\ee  
with  
\be 
\label{W}
L W(\phi)=-\frac{3}{2}-\frac{\phi^{2}}{2}-\frac{\phi^{4}}{4\phi_{M}^{2}}+\frac{\phi^6}{10}
\;,
\label{eq:superpotential}
\ee
where $\phi_{M}$ is the free parameter and $L$ is a length scale. The fact that the potential $V$ can be derived from a superpotential $W$ is often associated to supersymmetry. However, we stress that our choice is motivated solely by technical convenience; supersymmetry plays no role in our analysis. For any value of $\phi_{M}$, the dual field theory is a CFT deformed by a dimension-three operator with a constant source $\mm$, which introduces a characteristic mass scale in the QFT. The resulting properties of the theory---including its renormalization-group (RG) flow, correlation functions, and thermodynamics---are then fully determined by the form of the potential $V(\phi)$. Fig.~\ref{potpot} shows the potential for several values of $\phi_M$ that will be of interest below.

Extrema of the potential correspond to exact AdS solutions with constant $\phi$ and AdS radius $L$. In the dual field theory, these geometries represent RG fixed points with a number of degrees of freedom scaling as $N^2\sim L^3/G_5$. In top-down models this relation is known precisely. For example, in the case in which the gauge theory is $\mathcal N=4$ super Yang-Mills  with $N$ colours 
\begin{equation}
\label{defN}
\frac{L^3}{8\pi G_5}=\frac{N^2}{4\pi^2}\;.
\end{equation}
In our bottom-up model we will take this as a definition of the number of degrees of freedom in the QFT at the ultraviolet (UV) fixed point,   $\phi=0$. Henceforth we will set $L=1$. Extrema of the superpotential always correspond to extrema of  the potential, but the reverse is not generally true. In this paper we will restrict ourselves to values of $\phi_M$ for which both $V$ and $W$ have a maximum at $\phi=0$ and a minimum at some $\phi=\phi_E$, with no other extrema in between, as in Fig.~\ref{potpot}.

This holographic theory is a simple yet remarkably rich model.  Placed in flat space at finite temperature, the model exhibits a nontrivial phase structure~\cite{Bea:2018whf}: depending on the value of $\phi_M$, it undergoes phase transitions of varying orders and features regions of parameter space where a false vacuum appears.  
In this paper we will investigate the properties of this model when placed in a fixed dS background. 

\subsection{Scheme dependence}
\label{scheme}
In this section, we review the scheme dependence of a QFT in a curved background, which plays an important role in our analysis. We closely follow Refs.~\cite{Casalderrey-Solana:2020vls,Ecker:2021cvz}. We refer the reader to these references for additional details. 

Near the AdS boundary, the bulk metric can be written in the so-called Fefferman--Graham (FG) gauge 
\begin{equation}
\label{metricFG}
ds^2= \frac{dr^2}{4r^2}+\gamma_{\mu\nu}(r,x)d x^\mu d x^\nu \,. 
\end{equation}
The boundary is located at $r=0$ and is parametrised by the coordinates $x^\mu$ with \mbox{$\mu=0, \ldots, 3$}. Near the boundary the metric and the scalar behave as 
\begin{equation}
\label{seriesVac}
\gamma_{\mu\nu}(r,x) \sim \frac{g_{\mu\nu}(x)}{r} \sac \phi \sim M r^{1/2} \,,
\end{equation}
where $g_{\mu\nu}(x)$ is the boundary metric. 
Substituting this in the first term of the action \eqref{eq:Sgen} we see that it suffers from large-volume divergences.  These divergences can be regularised and renormalised by a procedure called holographic renormalisation (see e.g.~\cite{deHaro:2000vlm,Bianchi:2001de,Bianchi:2001kw}), which makes the action finite and the variational principle well-defined. 
This procedure is implemented  by including in \eqref{eq:Sgen} the counterterm action 
\beq
\label{eq:Sct}
 S_{ct} = \frac{1}{8\pi G_5} \int d^4 x\sqrt{-\gamma} \Bigg[  \left(
-\frac{1}{8}R-\frac{3}{2}-\frac{1}{2}\phi^2 \right) 
+ \frac{1}{2} \left( \log r\right)  \mathcal{A}+ 
  \Big(  \,\alpha \mathcal{A} + \beta\phi^4  \Big) \Bigg] \,,
\eeq
where $\alpha$ and $\beta$  are real constants. The terms in the final brackets correspond to finite counterterms; we have omitted additional  contributions of this type that are not relevant for the present discussion. The action is evaluated on a timelike  hypersurface at constant $r$ near the AdS boundary with induced metric 
$\gamma_{\mu\nu}$. In this and all subsequent equations in this section, metric-dependent quantities, such as the Ricci scalar $R$, are  constructed from the induced metric $\gamma_{\mu\nu}$.
 The second term of \eqref{eq:Sgen} is also understood to be evaluated on this slice, the first term  of \eqref{eq:Sgen} is understood to be evaluated by integrating down to this slice, and the limit $r \to 0$ is understood to be taken at the end of the calculation. 

In \eqref{eq:Sct},  $\mathcal{A}(\gamma_{\mu\nu}, \phi)$ is the so-called conformal anomaly,
\be
\mathcal{A}=\mathcal{A}_g + \mathcal{A}_\phi \,,
\ee
where
\be
\label{eq:Ag}
\mathcal{A}_g = \frac{1}{16}(R^{\mu\nu}R_{\mu\nu}-\frac{1}{3}R^2)
\ee
is the holographic gravitational conformal anomaly and 
\be
\label{eq:Aphi}
\mathcal{A}_{\phi}=-\frac{\phi^2}{12}R 
\ee
is the conformal anomaly due to matter.
In  dS space these become
\begin{eqnarray}
\label{often}
\mathcal{A}_g  &=& -\frac{3}{4} H^4 \,,\\[1mm]
\mathcal{A}_{\phi} &=& -M^2 H^2 \,. 
\end{eqnarray}
Flat space is obtained by simply setting $H=0$.

The freedom to add the terms in \eqref{eq:Sct} with arbitrary coefficients $\alpha$ and 
$\beta$ is part of the general freedom in the choice of renormalisation scheme. 
The value of $\alpha$ can be shifted by a scale transformation, which is implemented via the following rescaling of the  coordinates 
\begin{equation}
\label{rescresc}
x_\mu = \lambda x'_\mu \,, \qquad 
r = \lambda^2 r' \,,
\end{equation}
where $\lambda$ is a positive real number.  
In the QFT, this is equivalent to rescaling $H$ and $M$ through 
\begin{equation}\label{eq:rescale}
H' = \lambda H \,, \qquad M' = \lambda M \,.
\end{equation}
Under this transformation, the counterterm action shifts by  $(\log \lambda) \mathcal{A}$, which in turn can be absorbed through the redefinition $\alpha \to \alpha + \log \lambda$. The freedom to rescale $r$, or equivalently to shift $\alpha$, is thus the freedom to choose a renormalisation scale. We may therefore think of $\alpha$ as related to the renormalization group scale or subtraction point $\mu$ through
\beq
\label{rg}
\alpha \sim \log \mu \,.
\eeq
Reference to this arbitrary scale is necessary to define the stress tensor of the QFT. For example, under the transformation \eqref{rescresc}-\eqref{eq:rescale},
the QFT energy density extracted from the holographic renormalizaton procedure transforms as
\beq
\label{additive}
\rho(\lambda H, \lambda M)  = 
\lambda^4 \rho(H, M) + \lambda^4 \log \lambda \, 
\left( \frac{N^2}{4\pi^2}\right)\, H^2 M^2
  \,.
\eeq
This immediately implies that the energy density must take the form
\beq
\label{stress}
\rho ( H,  M)  = 
M^4 \, f \left( \frac{H}{M} \right) + \log \left( \frac{H}{\mu} \right) \, 
\left( \frac{N^2}{4\pi^2}\right) \, H^2 M^2\,,
\eeq
with $\mu$ is the renormalization scale above. Note that the logarithmic term is negative for $H<\mu$, which can render the energy density negative for certain values of 
$H$ in some renormalization schemes. 

The counterterm associated to $\beta$  plays a distinct role because it is the only one that does not vanish for a flat boundary metric. In this paper we will set  
\beq
\label{special}
\beta=- \frac{1}{4\phi_M^2} \,.
\eeq
With this choice,  the energy density of the QFT theory vanishes identically in flat space. 

Upon variation with respect to the boundary metric, the finite counterterms give contributions to the QFT stress tensor, which therefore we can write as 
\beq
\label{stst}
T_{\mu\nu}=T_{\mu\nu}^{(0)} + \alpha  \, T_{\mu\nu}^{(\alpha)}  + \beta \, T_{\mu\nu}^{(\beta)}    \,,
\eeq
where $T_{\mu\nu}^{(0)}$ denotes the stress tensor in the scheme 
$\alpha=\beta=0$. This means that, in the absence of dynamical gravity, the boundary stress tensor is ambiguous to the extent that the coefficients of the  counterterms are arbitrary. However, when the QFT is coupled to  dynamical  gravity, these coefficients simply renormalize the gravitational couplings and the ambiguity is replaced by the physical specification of the renormalized couplings \cite{Birrell:1982ix}. To see this, consider the four-dimensional Einstein equations  coupled to the QFT stress tensor \eqref{stst}: 
\beq
\label{EE}
\msp^2 \left( R_{\mu\nu} - \frac{1}{2} \, R g_{\mu\nu} +\Lambda g_{\mu\nu}\right)  = 
 N^{-2}\, T_{\mu\nu}\,.
\eeq
Note that this equation has a well defined large-$N$ limit provided $\msp$ is kept fixed, as expected from the discussion below \eqref{species}. 

To start, let us set $\alpha=0$ and  consider  the contribution to the  stress tensor of $T_{\mu\nu}^{(\beta)}$, which takes the form
\be
\label{Tbeta}
T_{\mu\nu}^{(\beta)} =  \frac{1}{8\pi G_5} \,M^4\, g_{\mu\nu} 
= \frac{N^2}{4\pi^2}\, M^4\, g_{\mu\nu} \,,
\ee
where we have made use of \eqref{defN}. Moving this term to the left-hand side of \eqref{EE} we can write Einstein equations in the form  
\be
\label{EE2}
\msp^2 \left( R_{\mu\nu} - \frac{1}{2} \, R g_{\mu\nu}\right) + \msp^2 \Lambda_{\rm{ren}} \, g_{\mu\nu} =  N^{-2}\, T_{\mu\nu}^{(0)} \,,
\ee
with 
\be
\label{lambdaren}
 \Lambda_{\rm{ren}} =  \Lambda - 
\frac{M^4}{4\pi^2 \msp^2} \, \beta  \,.
\ee
Note that this equation is $N$-independent. We see that the effect of the $\beta$-counterterm is simply to renormalize the bare cosmological constant 
$\Lambda$ in Einstein equations. In other words, $\Lambda$ and $\beta$ are not separately meaningful, only the combination  $\Lambda_{\rm{ren}}$ is. It is therefore convenient to choose $\beta$ as in \eqref{special}, since in this case flat space is a solution of \eqref{EE} with $\Lambda=0$.  

Consider now adding the contribution of the $\alpha$-counterterm and assume that the boundary metric is of FRWL type, which includes the dS case of interest. Under these circumstances 
\beq
T_{\mu\nu}^{(\alpha)} = - \frac{N^2}{4\pi^2} \frac{M^2}{3} \left( R_{\mu\nu} - \frac{1}{2}
R g_{\mu\nu} \right) \,.
\eeq
Moving this term to the left-hand side of EE, these become 
\beq
\label{EEren}
\mspren^2 \left( R_{\mu\nu} - \frac{1}{2} \, R g_{\mu\nu} +\Lambda_{\rm{ren}} g_{\mu\nu}\right)  = 
 N^{-2}\, T_{\mu\nu}^{(0)}\,,
\eeq
with 
\beq
\label{renren}
\mspren^2 = \msp^2 +\frac{M^2}{12 \pi^2}\, \alpha\,.
\eeq
We see that the effect of the $\alpha$-counterterms is simply to renormalize the bare, four-dimensional quantum gravity scale $\msp$. In other words, $\msp$ and $\alpha$ are not separately meaningful, only the combination $\mspren$ is. Since  $\alpha$ is associated to renormalization group transformations, Eq.~\eqref{renren} can be seen as the renormalization group equation for the running of the species scale. For convenience we will work in the scheme 
\beq
\label{aa}
\alpha=0 \,.
\eeq

\subsection{\label{sec:dSbg}de Sitter}  
When matter is placed in dS apace, its energy density dilutes as the Universe expands. The effects of this expansion have already been studied in the context of holography in~\cite{Buchel:2017pto,Buchel:2019qcq,Casalderrey-Solana:2020vls}. These works have shown that, at sufficiently late times, the field theory asymptotically approaches a state with a simple time dependence and dS-invariance at all scales. In~\cite{Buchel:2017pto}, this state was identified as the Bunch-Davies vacuum of the field theory. 
As noted above, the holographic models of Refs.~\cite{Buchel:2017pto,Buchel:2019qcq,Casalderrey-Solana:2020vls} do not display the multiple branches illustrated in Fig.~\ref{Edens059_D}. In this work, we focus on characterizing the late-time, dS-invariant states that arise when this branch structure is present.

Although our results will be independent of the choice of dS coordinates, for concreteness we work in the flat  slicing, in which the metric takes the form
\be
ds^2_{dS_4}=-dt^2 + e^{2Ht} dx^2\;,
\ee
with $H$ the Hubble rate. The late-time state of the QFT in de Sitter is described in the dual theory by a solution of the five-dimensional action~\eqref{eq:Sgen} which, 
following~\cite{Buchel:2017pto,Buchel:2019qcq,Casalderrey-Solana:2020vls}, may be written in Eddington--Finkelstein (EF) type coordinates as 
\begin{equation}
\label{eq:EFantsatz}
ds^2=-A(u)dt^2 + \Sigma(u)^2 e^{2 H t} dx^2 -\frac{2}{u^2} dt du\;,
\end{equation} 
with $u$ a holographic coordinate such that the boundary is located at $u=0$. Note that we use the same label for the bulk and boundary times, since they coincide at the boundary.  As we will see, our assumption that the metric functions $A$ and $\Sigma$ are time-independent  guarantees that the dual QFT states are dS-invariant.      The use of horizon-penetrating coordinates in the bulk will be essential for our purposes, as they will enable us to probe features such as the apparent horizon, which lies behind the event horizon.  Inserting the  ansatz~\eqref{eq:EFantsatz} in the Einstein-scalar equations, we obtain
\begin{eqnarray}
\label{ets1}
0&=&-\frac{6 \Sigma'}{\Sigma u}-\frac{3 \Sigma''}{\Sigma}-2 \left(\phi '\right)^2 \,,
\\
\label{ets2}
0&=& u^2 \phi ' \left(u^2 A'+A u \left(\frac{3 u \Sigma'}{\Sigma}+2\right)-3 H\right)+A u^4 \phi
   ''-V' \,,
   \\
   \label{ets3}
   0&=&
 -\frac{3 u^2 A' \Sigma'}{2 \Sigma}+A \left(-\frac{3 u^2 \left(\Sigma'\right)^2}{\Sigma^2}-\frac{6 u
   \Sigma'}{\Sigma}-u^2 \left(\phi '\right)^2\right)-\frac{3 A u^2 \Sigma''}{\Sigma}+\frac{9 H \Sigma'}{\Sigma}-\frac{2
   V}{u^2}\,,
   \\
   \label{ets4}
   0&=&\frac{1}{2} \Sigma^2 u^4 A''+A u^4 \left(\Sigma'\right)^2+2 A \Sigma
   u^4 \Sigma'' + 
   \\
   &&
   +2 \Sigma u^2 \Sigma' \left(u^2 A'+2 A u-3 H\right)+\frac{1}{2} \Sigma^2 \left(2 u^3 A'+2 A u^4
   \left(\phi '\right)^2+4 V\right) \,, \nonumber
   \\
   \label{ets5}
   0&=&
   -\frac{3 A^2 u^4
   \left(\Sigma'\right)^2}{\Sigma^2}-\frac{3 A^2 u^4 \Sigma''}{\Sigma}-
   \\
   &&
   -\frac{3 A u^2 \Sigma' \left(u^2 A'+4 A u-8 H\right)}{2 \Sigma}+\frac{1}{2} \left(-3 H \left(u^2
   A'+2 H\right)-2 A^2 u^4 \left(\phi '\right)^2-4 A V\right)\,,
\nonumber
\end{eqnarray}
where $'$ denotes differentiation with respect to $u$. 
The requirement that the boundary metric is dS$_4$ imposes that, in the vicinity of $u=0$, these functions behave as
\begin{eqnarray}
\label{phinb}
\phi&=&\mm  u - \mm  \xi  u^2+ u^3 \left(H^2 \mm  \log (u)+\phi _2\right) +O(u)^4 \,,\\
\label{Anb}
A&=&
\frac{1}{u^2}+\frac{2 \xi }{ u}
-H^2-\frac{2 \mm ^2}{3}+\xi
   ^2+ u^2 a_4  -\frac{2}{3} \mm^2 H^2 u^2 \log(u) 
 +O(u)^3 \,,
\\
\label{Sbcb}
\Sigma &=&\frac{1}{u}+(H+\xi )-\frac{\mm ^2 u}{3}+\frac{1}{9} u^2 \left(3 \mm ^2 \xi -H
   \mm ^2\right)+
\\
&&+\frac{1}{108} u^3 \left(9 H^2 \mm ^2-36 H^2 \mm ^2 \log
   (u)+24 H \mm ^2 \xi +2 \mm ^4-36 \mm  \phi _2\right)+O\left(u^4\right) \,.
   \nonumber
\end{eqnarray}
Here $\mm$ is the source of the 
dimension-three operator that breaks conformal invariance in the dual field theory and $\phi_2$ controls the expectation value of this operator. The gauge parameter $\xi$ is associated to the residual reparametrization invariance  $1/u \rightarrow 1/u+ c$. This leaves the form of the metric~\eqref{eq:EFantsatz} invariant but 
induces induces the shifts 
\beq
\label{shishi}
\xi\rightarrow \xi+ c \,,\qquad 
\phi_2 \rightarrow \phi_2 + M c^2 + 2 M c \, \xi
\,.
\eeq
For the dS-invariant solution,   
$a_4$ is given by 
\be
\label{eq:a4dS}
a_4=
\frac{1}{54} \mm ^2 \left(21 H^2+36 \xi ^2\right)+\frac{4 \mm
   ^4}{27}-\frac{2 \mm  \phi _2}{3}\;, 
\ee
which is invariant under the shifts \eqref{shishi}.

Not all equations~\eqref{ets1}-\eqref{ets5} are independent. In fact, a simple manipulation leads to the following equation for $\Sigma$:
\be
\label{eqtoSformal}
H \left(-\frac{A'}{2 A}-\frac{H}{A u^2}+\frac{\Sigma'}{\Sigma}\right)=0\;.
\ee
If  $H\neq 0$ the  solution is 
\be
\label{SigmaFormal}
\Sigma=\sqrt{A(u)} \exp
\left(H \int_0^u \frac{1}{y^2 A(y)} \, dy \right)\;,
\ee
where we have also imposed the boundary condition~\eqref{Sbcb}.  This formal solution allows us to reduce the equations~\eqref{ets1}-\eqref{ets5} to only two independent equations:
\begin{eqnarray}
\label{etsr1}
 0&=&A''+\frac{2 A'}{u}-\frac{\left(A'\right)^2}{2 A}+\frac{2 H^2}{A u^4}+\frac{4}{3} A
   \left(\phi '\right)^2  \,,
   \\
   \label{etsr2}
0&=&\phi '' + \frac{\phi ' \left(\frac{5 u^4 A'}{2}+2 A u^3\right)}{A u^4}-\frac{V'(\phi )}{A
   u^4}\;.
\end{eqnarray}

We are interested in solutions to these equations that admit a bulk event horizon. In this gauge, the location $u_h$ of the event horizon is the solution of the  condition~\cite{Casalderrey-Solana:2020vls}
\beq
\label{EHcondition}
A(u_h)=0 
\,.
\eeq
By exploiting the reparameterization invariance of our coordinate choice, we can set \mbox{$u_h=1$}. Near the horizon, the equations admit a power-series solution whose first few terms are
\begin{eqnarray}
\label{Anh}
A(u) &=&2 H (1-u)+(1-u)^2 \left(2 H-\frac{V(\phi_h)}{3}\right) +\dots\,,
\\
\label{phinh}
\phi(u)&=&\phi_h-\frac{(u-1) V'(\phi_h)}{5 H} + \dots\;,
\end{eqnarray}
where $\phi_h$ is the value of the scalar field at the horizon. These expressions serve to numerically construct gravitational solutions of the form given by equation~\eqref{eq:EFantsatz}.

All the solutions constructed in this way are dual to dS-symmetric states in the dual field theory. This is a consequence of the time independence of the $A$ and $\Sigma$ functions, which imposes the relation~\eqref{SigmaFormal} between them. This relation
implies that, after a change of variables, the metric~\eqref{eq:EFantsatz} can be put in the Fefferman--Graham-like form 
\be
\label{eq:dwmetric}
ds^2=\frac{dz^2}{z^2} + A(z)  ds^2_{\mathrm{dS}_4}\;.
\ee
 Indeed, if $\tau$ is the time coordinate in the flat slicing of the dS$_4$ metric in~\eqref{eq:dwmetric}, then the following change of coordinates brings the EF metric~\eqref{eq:EFantsatz} to the FG form~\eqref{eq:dwmetric}: 
\be
 \frac{z'(u)}{z(u)} = 
 \frac{1}{u^2 \sqrt{A(u)}}\;,  \quad  \quad \tau=t + \int_0^u \frac{d \tilde u }{\tilde u^2 A(\tilde u)}\;.
 \ee 
The form of the metric~\eqref{eq:dwmetric} makes the full dS$_4$ symmetry of the solutions manifest. The factorization implied by the second term on the right-hand side of~\eqref{eq:dwmetric} is possible because dS is maximally symmetric and, as a consequence, its Ricci tensor is proportional to its metric with a constant coefficient. Relatedly, inspection of the Einstein-scalar equations in the gauge~\eqref{eq:dwmetric} reveals that the functions $A, \Sigma$ and $\phi$ are independent of the choice of coordinates in dS$_4$. 
 
To determine all dS--invariant states, we work in units where $H=1$ and vary the mass parameter $M$, since $H$ appears explicitly in the equations of motion. We then compute the QFT stress tensor using the renormalization scheme defined in \eqref{special} and \eqref{aa}, and plot the dimensionless ratio $\mathcal{E}/M^{4}$. 
We emphasize that this procedure is not equivalent to working units in which $M=1$, although the two choices are related. Indeed, transforming from units with $H=1$ to those with $M=1$ requires rescaling both $H$ and $M$ by a factor $\lambda = 1/M$, as in \eqref{eq:rescale}. Under this rescaling, the energy density does not transform by a simple multiplicative factor; rather, it acquires an additive contribution proportional to $\log(H/M)$, originating from the conformal anomaly, as given in \eqref{additive}.

We generate the solutions by numerically integrating  equations~\eqref{ets1}-\eqref{ets2} from the horizon. After specifying a value of $\phi_h$ we use the expansion in equations~\eqref{Anh} and~\eqref {phinh} to set the initial conditions for integration. We fit the numerically generated fields to their asymptotic form, given by equations~\eqref{phinb} and~\eqref{Anb}, to extract the value of the source $\mm$ and the coefficients $\xi$ and $\phi_2$. Since our problem only involves two dimensionful parameters, the values for the source we extracted from this procedure correspond to a value of the ratio $\mm/H$. Therefore, by scanning over $\phi_h$, we obtain different values of $\mm/H$.

As shown in~\cite{Bea:2018whf}, the family of models we study have different branches, associated to different types of Renormalization Group (RG) flows. In this paper we focus on those RG-flows that in the UV flow to the undeformed CFT, with $\mm=0$. This implies that we explore the region $0<\phi_h<\phi_E$, with $\phi_E$ the minimum of both the potential~\eqref{V} and of the superpotential~\eqref{W}, which depends on the value of $\phi_M$. Since the gauge theory vacuum in flat space corresponds to a  RG flow from $\phi=0$ to $\phi=\phi_E$, the limit $\phi_h\rightarrow \phi_E$ corresponds to $H/\mm \rightarrow 0$.

\subsection{\label{sec:FTgen}Finite temperature}
For completeness, and to facilitate the comparison between the finite-temperature and finite-curvature states of the field theory, we briefly review the construction of thermal solutions within this holographic model. One key difference between the two cases is that the dS solutions are invariant under a larger symmetry group  than the thermal solutions, and are consequently more constrained. 

The gravitational duals of finite-temperature states can be found in the EF ansatz:
\begin{equation}
\label{eq:EFantsatzT}
ds^2=-A(u)dt^2 + \Sigma(u)^2 dx^2 -\frac{2}{u^2} dt du\;,
\end{equation} 
which coincides with equation~\eqref{eq:EFantsatz} in the limit $H=0$. As a consequence, the equations of motion and the boundary conditions that these metric functions satisfy are given by the $H=0$ limit of equations~\eqref{ets1}--\eqref{ets5} and~\eqref{phinb}--\eqref{Sbcb}, respectively. However, since $H=0$,  equation~\eqref{eqtoSformal} is trivially solved and hence $\Sigma$ is not  given by~\eqref{SigmaFormal}. As a result, the coefficient \( a_4 \) is no longer fixed by boundary data through equation~\eqref{eq:a4dS}. This means that the near-boundary expansion is now characterized by two independent coefficients, \( a_4 \) and \( \phi_2 \), which together determine the expectation values of the stress tensor and the dimension-three operator that deforms the UV CFT.

A  solution dual  to a thermal state also possesses an event horizon whose location is given by the solution of~\eqref{EHcondition}, and we can again set $u_h=1$, as in the finite-curvature case. Close to the horizon, we can also find a power series solution whose first few orders are
\begin{eqnarray}
       \label{eq:phinhft}
    \phi&=&\phi _h-\frac{(u-1) V'\left(\phi _h\right)}{A_1}+\frac{(u-1)^2 V'\left(\phi _h\right) \left(12 A_1+3 V''\left(\phi
   _h\right)+8 V\left(\phi _h\right)\right)}{12 A_1^2}+\dots \,,\qquad
   \\
          \label{eq:Anhft}
   A&=&A_1 (1-u)+\frac{1}{3} (1-u)^2 \left(3 A_1+2 V\left(\phi _h\right)\right)+\dots\,,
   \\[2mm]
          \label{eq:Snhft}
   \Sigma&=&
   \Sigma _{EH}-\frac{4 \Sigma _{EH} (1-u) V\left(\phi _h\right)}{3 A_1}-\frac{(1-u)^2 \left(4 A_1 \Sigma_{EH} V\left(\phi
   _h\right)+\Sigma_{EH} V'\left(\phi _h\right){}^2\right)}{3 A_1^2}\,.
\end{eqnarray}
As in the dS case, $\phi_h$ is the value of the scalar field at the horizon. Unlike the dS case, however, the near-horizon behavior is not locally fully determined once $\phi_h$ is specified, since solutions can be found for arbitrary constants $A_1$ and $\Sigma_{EH}$. Nevertheless, demanding that the solution matches the boundary behavior given by equations~\eqref{phinb}--\eqref{Sbcb} fixes  the values of these constants.

The strategy to find these thermal solutions is similar to that in the dS case. After specifying the value of $\phi_h$, we use equations~\eqref{eq:phinhft}--\eqref{eq:Snhft} with arbitrary values of $A_1$ and $\Sigma_{EH}$ as initial conditions to solve the equations of motion. After an appropriate rescaling of the coordinates, the numerically generated solution can be put in the form of~\eqref{phinb}--\eqref{Sbcb} for a fixed value of $\mm$.\footnote{In practice, we follow the algorithm spelled out in~\cite{DeWolfe:2010he}, which uses a different gauge.} From this solution, we can extract the temperature 
of the field theory, given by the surface gravity
of the dual black brane:
\be
\frac{T}{\mm}=  \frac{A_1}{4\pi}\;.
\ee
In conclusion, we find that, as in the dS case, $\phi_h$ provides a one-to-one parametrization of all thermal solutions.

\subsection{Stress tensor}
As expleined in Section \ref{scheme}, the expectation value of the stress tensor for the field theory states dual to our gravitational solutions can  be extracted holographically.
Following~\cite{Bianchi:2001de,Bianchi:2001kw}, the authors of~\cite{Casalderrey-Solana:2020vls} analyzed the stress tensor for arbitrary homogeneous states of this holographic model in dS space in the flat slicing. Due to the assumed symmetries, only the diagonal components of the expectation value are non-vanishing. Factoring out an overall normalization factor, 
\begin{equation}
\left<T^\mu_{\ \nu} \right>\equiv \frac{N^2}{2\pi^2} \, \mathrm{diag} \left(-\mathcal{E}, \mathcal{P}, \mathcal{P}, \mathcal{P} \right)\;,
\end{equation}
the energy density and the pressure are given by
\begin{eqnarray}
 \label{eq:Ehol}
    \mathcal{E}&=&-\frac{3 a_4}{4}-\frac{1}{2} \alpha  H^2 \mm^2+\mm^2 \left(\frac{19 H^2}{24}+\xi
   ^2\right)+\frac{3 H^4}{16}-\left(\beta -\frac{7}{36}\right) \mm^4-\mm \phi _2\;,
   \,\,\,\,\,\,\,\,
   \\[2mm]
    \label{eq:Phol}
   \mathcal{P}&=&-\frac{a_4}{4}+\frac{1}{2} \alpha  H^2 \mm^2-\frac{1}{3} \mm^2 \left(\frac{29 H^2}{24}+\xi
   ^2\right)-\frac{3 H^4}{16}+\left(\beta -\frac{5}{108}\right) \mm^4+\frac{\mm \phi _2}{3}\;.\,\,\,\,\,\,\,\,
\end{eqnarray}
Here he have displayed the scheme parameters $\alpha$ and $\beta$. Below we will set them to the values \eqref{aa} and \eqref{special}. These expressions apply to holographic solutions dual to dS-symmetric states as well as to thermal solutions in flat space, which are  obtained by setting \( H = 0 \). In the dS case, the coefficients \( a_4 \) and \( \phi_2 \) are not independent but are related through equation~\eqref{eq:a4dS}. Imposing this relation leads to the condition
\begin{equation}
\label{eos1}
\mathcal{E} = -\mathcal{P}\;,
\end{equation}
which corresponds to an equation of state 
\beq
\label{eos2}
 w = \frac{\mathcal{P}}{\mathcal{E}} = -1 \,. 
 \eeq
This  is a direct consequence of the maximal symmetry of dS, which implies that the stress tensor must be proportional to the metric,
\begin{equation}
\label{epsi}
\left<T_{\mu \nu}\right> = \rho(H, \mm)\, g_{\mu \nu}\;,
\end{equation}
where the proportionality factor is given by 
\begin{equation}
    \rho(H, \mm) = \frac{N^2}{2\pi^2} \mathcal{E} \,.
\end{equation}

We emphasize that the nature of \eqref{eos2} is physically distinct from that of the thermodynamic equation of state governing an adiabatically expanding fluid. The latter relies on the assumption of local thermal equilibrium, whereas \eqref{eos2} is entirely fixed by symmetry. A useful analogy is provided by a CFT, in which conformal invariance enforces the relation $\mathcal{E}=3\bar{\mathcal{P}}$ even in states that are far from equilibrium, where $\bar{\mathcal{P}}$ denotes the average pressure.

\subsection{Free energy}
\label{free-energy}
As we show below, for suitable parameter choices our  model admits multiple dS-invariant solutions with the same value of $H$, as in Ref.~\cite{Ghosh:2017big}. As in the thermal case, each of these solutions corresponds to a saddle point of the corresponding path integral. For thermal states in flat space, the thermodynamically preferred solution is the one with the lowest free energy $F$, or equivalently, the one with the lowest Euclidean action $S_E=F/T$. For homogeneous states, this criterion reduces to the minimization of the corresponding densities, related through $s_E=f/T$. Following \cite{Ghosh:2017big}, we will adopt the viewpoint that an analogous criterion applies in dS. We emphasize, however, that this interpretation does not affect the existence of self-sustained dS solutions---although it may have implications for their stability, a point to which we will return below.

For the metric \eqref{eq:dwmetric} the classical action factorizes into two contributions: one depending only on the dS$_4$ coordinates, and another depending on the holographic coordinate $z$ and the curvature scale $H$. This factorization allows continuation to Euclidean space. If the dS$_4$ metric is written in global coordinates, then   the problem may be understood as determining the preferred spherically symmetric state of the Euclidean QFT  on the four-sphere. Unlike the thermal case, however, the Euclidean action in this context is not fixed by thermodynamic relations and must be computed explicitly.

After some manipulations, the action~\eqref{eq:Sgen} evaluated on a dS-symmetric solution of the form \eqref{eq:dwmetric} is given by
\be
S_E=-\frac{8 \pi^2}{3 H^4} f_H\;,
\ee
where $f_H$ is defined by the integral
\be
\label{eq:fHdef}
f_H=\lim_{\epsilon \rightarrow 0}\left( \int_\epsilon^{z_H} dz  \frac{3 H^2}{2 z} A(z)  \, + f_{ct} (\epsilon) \right)\,,  
\ee
where $z_H$ and $\epsilon$ are the horizon position and a  near boundary regulator in the coordinates \eqref{eq:dwmetric}, respectively.  The term $f_{ct} (\epsilon)$
is given by 
\begin{eqnarray}
f_{ct} (\epsilon) &=&-\frac{3}{4} \epsilon A(\epsilon) A'(\epsilon)
   -\frac{1}{2} A(\epsilon)
   \left(A(\epsilon) \left(\phi (\epsilon)^2+3\right)+3 H^2\right)
-
     \\
   &&
-\frac{1}{8} H^2 \log \left(\epsilon^2 \right) \left(4 A(\epsilon) \phi (\epsilon)^2+3
   H^2\right)
   +\beta  A(\epsilon)^2 \phi (\epsilon)^4
      +\alpha  \left(-H^2 A(\epsilon) \phi (\epsilon)^2-\frac{3 H^4}{4}\right).    \nonumber
\end{eqnarray} 
The first term in this expression arises from a total derivative contribution to the bulk action~\eqref{eq:Sgen}; the remaining terms originate from the counterterm action~\eqref{eq:Sct}. The expression is in general divergent as $\epsilon\to 0$, as can be seen by inserting the near-boundary expansions~\eqref{phinb}--\eqref{Sbcb} for the different fields. These divergences cancel precisely against the divergences of the integral appearing in equation~\eqref{eq:fHdef}, rendering $f_H$ finite. This cancellation is a consequence of holographic renormalization; nevertheless, it also induces an ambiguity in the definition of $f_H$, which depends explicitly on the scheme parameters $\alpha$ and $\beta$. Although the Euclidean action is scheme-dependent, the action  difference  between two dS-symmetric states with  the same expansion rate is not. This difference therefore provides an unambiguous criterion for selecting the preferred state.

\subsection{Entropy}
\label{entropy-section}
We have seen that the holographic duals of dS-symmetric states and thermal states  share many common features. Both are described by similar metric functions that satisfy closely related equations of motion. More importantly, in both cases the dual geometries possess an event horizon, and different states are uniquely characterized by the values of the bulk fields at the horizon.

The presence of a horizon plays a central role in the holographic description of thermal systems. In that context, the Bekenstein--Hawking entropy of the black brane coincides with the thermodynamic entropy of the QFT. Accordingly, the entropy density of the latter, $s$, is determined by the geometry of the event horizon through 
\be
\label{ss}
\frac{s}{\mm^3}\,\, =\,\, \frac{2\pi}{8\pi G_5}\frac{\Sigma^3_{EH}}{\mm^3} 
\,\,\equiv \,\, \frac{N^2}{2\pi^2}\frac{\mathcal{S}}{M^3}\;,
\ee
where $\Sigma_{EH}$ is defined in equation~\eqref{eq:Snhft}.

dS-symmetric solutions also possess a horizon. In particular, the spatial warp factor $\Sigma$ remains finite at the event horizon, just as in the thermal case. This observation suggests that a notion of entropy may likewise be associated with dS-symmetric states. However, there is an important difference from  the thermal case. As emphasized in~\cite{Buchel:2017pto,Buchel:2019qcq},  the geometries dual to de Sitter–symmetric states are explicitly time-dependent, in contrast to the static black-brane geometries dual to thermal states. Under these circumstances, the event and apparent horizons do not, in general, coincide. Studies of holographic dynamical systems indicate that, in time-dependent settings, it is the apparent horizon---not the event horizon---that encodes the entropy of the dual field theory \cite{Booth:2005qc,Figueras:2009iu,Engelhardt:2017aux,misc}. Although the apparent horizon is in general slice-dependent, the holographic framework often provides a preferred choice of slicing obtained by shooting ingoing null geodesics from the boundary. This prescription is motivated by the fluid/gravity correspondence \cite{Bhattacharyya:2007vjd} and, in the hydrodynamic regime, can be shown to reproduce the entropy production of the dual QFT. In our case, the preferred slicing is encoded in the EF form of the metric \eqref{eq:EFantsatz}. In more general situations, different choices of bulk slicing correspond to different coarse-grainings in the dual QFT \cite{Engelhardt:2017aux}, which are necessary to define an entropy that obeys the second law.

Following this line of ideas, \cite{Buchel:2017pto,Buchel:2019qcq}  
argued that the area of the apparent horizon in our context can be used to assign an entropy density to dS-symmetric states. For the metric \eqref{eq:EFantsatz}, the location of the apparent horizon is given by the solution of the condition~\cite{Casalderrey-Solana:2020vls}
\be
\label{eq:AHcond}
H \Sigma(u_{\rm AH}) - \frac{1}{2} u^2_{\rm AH} A(u_{\rm AH}) \Sigma'(u_{\rm AH}) = 0
 \;,
\ee
which, using Eq.~\eqref{SigmaFormal}, translates into
\beq
u^2_{\rm AH} A'(u_{\rm AH}) = 2H \,.
\label{AHcondition}
\eeq
Since outside the event horizon $A'(u)<0$, this condition implies that the apparent horizon must lie inside the event horizon, as expected on general grounds. 
To determine $u_{\rm AH}$, we numerically integrate the equations of motion inward from a point just inside the event horizon, continuing toward the bulk interior until condition~\eqref{eq:AHcond} is satisfied.  At this point we evaluate the spatial warp factor $\Sigma_{\rm AH} \equiv \Sigma(u_{\rm AH})$. After factoring out $N^2/2\pi^2$, as in \eqref{ss}, the entropy density per unit physical volume in the QFT is then
\be
\label{conscons}
\mathcal{S}_H = \pi\Sigma_{\rm AH}^3\;.
\ee
Since $\mathcal{S}_H$ is constant and the physical volume grows exponentially in time, the comoving entropy density increases as
\beq
\label{coco}
\mathcal{S}_{\textrm{com}}=\mathcal{S}_H \times e^{3Ht} \,. 
\eeq
In situations with multiple solutions for the same $H$, as in Fig.~\ref{Edens059_D}, the value of $\mathcal{S}_H$ depends on the branch under consideration.

This continuous entropy production reflects the out-of-equilibrium character of the QFT state. To clarify this point, it is useful to compare with a fluid in an FLRW universe with scale factor $a(t)$. In the absence of entropy production, the entropy density per unit physical volume scales as $\mathcal{S}_{\mathrm{phys}} \sim a^{-3}$. Since the physical volume grows as $a^{3}$, the comoving entropy density,
$\mathcal{S}_{\mathrm{com}} = a^{3}\mathcal{S}_{\mathrm{phys}}$, remains constant, corresponding to a constant total entropy.
If the fluid departs slightly from local thermal equilibrium, dissipation due to viscous effects leads to entropy production. In this case, $\mathcal{S}_{\mathrm{phys}}$ still decreases with time, but more slowly than $a^{-3}$. Consequently, the comoving entropy density increases slowly, reflecting a mild departure from equilibrium.
As we will argue in Section~\ref{sec:cosmo}, in our model the universe may approach a dS-invariant state at late times, once the fluid energy density has sufficiently diluted. In this regime, the departure from equilibrium is large: the physical entropy density no longer decreases, but instead approaches the constant value given in \eqref{conscons}. As a result, the comoving entropy density grows exponentially with time, as in \eqref{coco}. Remarkably, this behavior does not break any dS symmetries, since it is entirely driven by the growth of the physical volume. Equivalently, because the entropy density per unit physical volume remains constant, it does not single out a preferred frame and therefore preserves all dS symmetries.

We emphasize that, although the microscopic interpretation of the entropy \eqref{conscons} is not fully understood \cite{Buchel:2019qcq}, it is clear that it does not represent the Bekenstein–Hawking entropy of the cosmological dS horizon, since gravity on the boundary is non-dynamical. Instead, it captures properties of the quantum fields evolving in a fixed dS geometry. We will return to this point below.

\section{Phase structure}
\label{sec:fases}
In the previous section we reviewed the main tools used to characterize both dS-symmetric and thermal states of the  strongly coupled matter described by our holographic model. In this section, we use these tools to explore the phase structure of the model in more detail.

\subsection{Phases at small $H$}
\label{curvature-driven-phases}
We begin by examining the specific model of our family \eqref{eq:superpotential} with $\phi_M=0.59$. This choice provides a representative example for illustrating the novel dynamical phenomena that emerge in the regime  $H \ll \mm$.

\begin{figure}[t]
    \begin{center}
    	\includegraphics[width=0.60\textwidth]{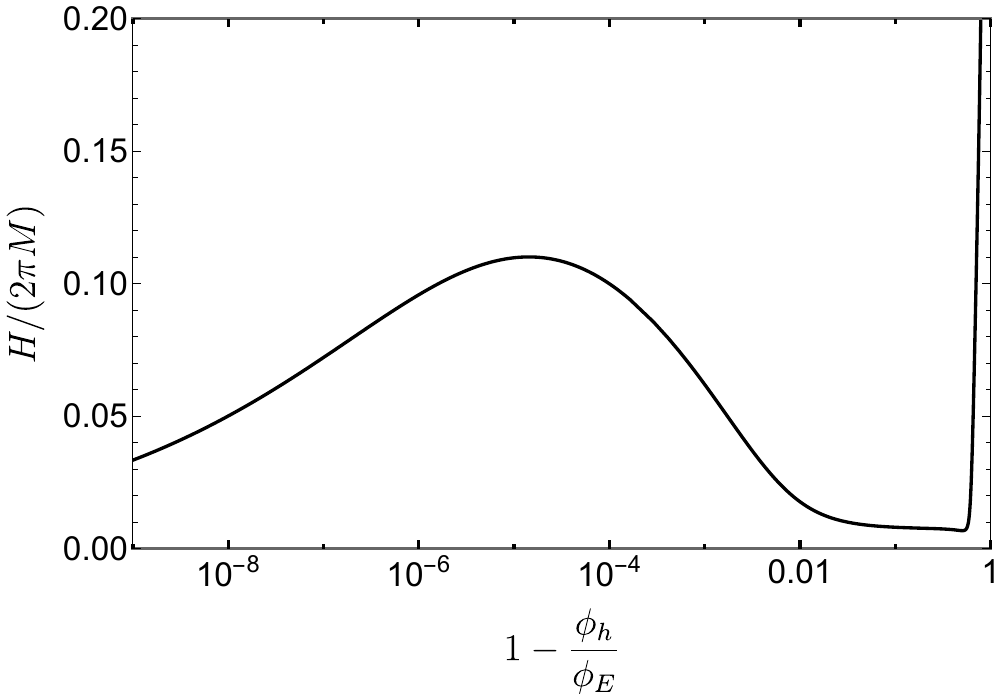}
    \end{center}
    \caption{\small\label{Hphihphiv059} 
    Relation between the Hubble rate of the QFT in dS and the value of the scalar field at the horizon in the dual bulk  for the model with  $\phi_M=0.59$. Multiple states coexist in the range of $H$ between the maximum and the minimum. $\phi_E$ corresponds to the first positive minimum of the potential $V(\phi)$.}
\end{figure}

As already anticipated, multiple bulk geometries may exist for certain values of the expansion rate $H$. Figure~\ref{Hphihphiv059} shows the relation between $H$ and the value of the scalar field at the horizon, $\phi_h$. 
This relation exhibits both a maximum and a minimum. Since $\phi_h$ provides a one-to-one parametrization of solutions, the presence of these extrema implies that, within this range, three distinct dS-symmetric states  exist for the same  expansion rate. This, in turn, suggests the possibility of curvature-driven quantum phase transitions between them, as discussed in~\cite{Ghosh:2017big}. The novel feature of our model is that the local minimum of the expansion rate occurs at a value much smaller than the characteristic QFT scale,  $\mm$. This implies the existence of multiple dS-symmetric states even when $H \ll \mm$. As we will see, this would play an important  role in the expansion history of a universe filled with this type of matter. In what follows, we explore several key properties that distinguish these coexisting states.

\begin{figure}[t!]
    \begin{center}
    \includegraphics[width=0.47\textwidth]{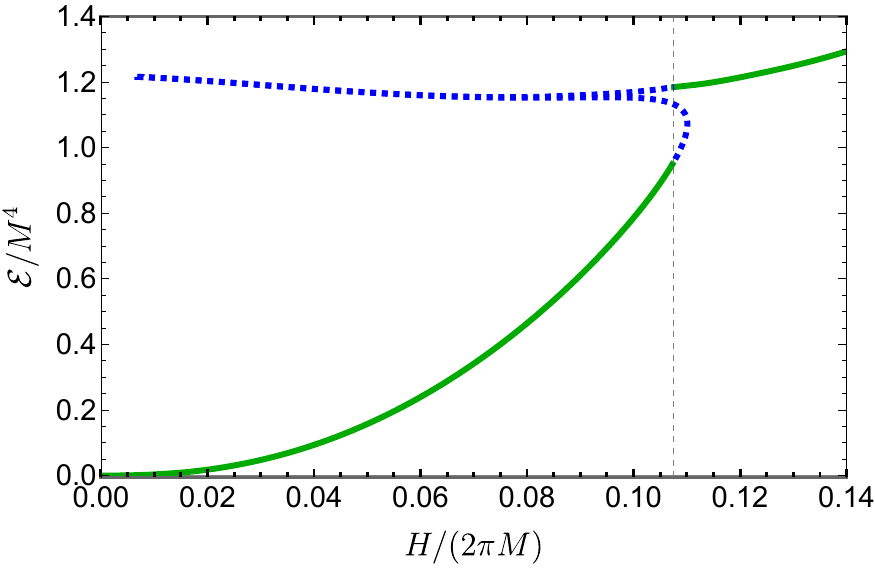}
    \hspace{5mm}
        \includegraphics[width=0.47\textwidth]{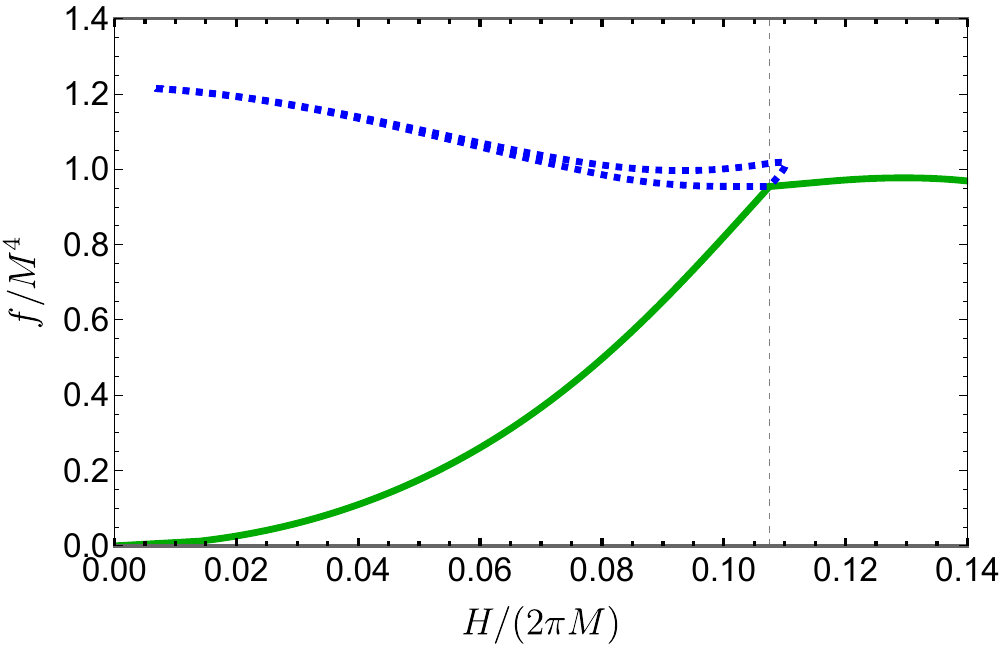}
    \end{center}
    \caption{\small\label{Edens059}
    Energy  (left) and free energy (right) densities as functions of the Hubble rate for the model with $\phi_M=0.59$. The plots exhibit multiple states separated by a finite energy-density gap down to small values of $H$. The branches of the phase transition are shown in solid green for globally stable states and dashed blue for the remaining states. The dashed vertical line corresponds to the critical de Sitter temperature of the phase transition. In our renormalization scheme, the energy density at small $H$ in this and similar  plots is slightly negative---see the comment below \eqref{stress}.}
\end{figure}

In Fig.~\ref{Edens059}(left), we reproduce Fig.~\ref{Edens059_D}(left) to facilitate comparison with the free energy density of the dS-symmetric states in this model, shown in Fig.~\ref{Edens059}(right).
As discussed above, these quantities depend on the choice of renormalization scheme. The results shown here correspond to our choice of scheme discussed above.  Remarkably, even at very small expansion rates, the difference between the highest and lowest energy densities remains finite and set by the characteristic scale of the theory, $\mm^4$. Importantly, while the absolute value of the energy density in each branch is scheme-dependent, their difference is not, making it a genuine physical feature of the model.

While multiple dS–symmetric states can coexist at the same expansion rate, they are not necessarily stable. Assessing the full dynamical stability of each state is a complex problem beyond the scope of this work. Determining their local stability  would require a detailed analysis of linear perturbations  along the lines of \cite{Buchel:2017lhu,Mashayekhi:2025jyg}, which we leave for future investigation. 

Building on the thermodynamic analogy discussed in Section~\ref{free-energy}, one may use the free energy density as a criterion for global stability: the globally preferred state is the one with the lowest free energy, to which the other solutions are expected to eventually decay. The free energy density, shown in Figure~\ref{Edens059}(right), exhibits the same qualitative behavior as the familiar swallow-tail structure of first-order thermal phase transitions: it is continuous but multivalued within a certain range of expansion rates. 
The stable solution is shown as a solid green curve, while the non-stable ones are indicated by dashed blue curves. In this way, we identify a phase transition where the free energy of the stable state develops a discontinuous derivative at $T_\textrm{dS}/\mm \simeq 0.107$, with $T_\textrm{dS}\equiv H/2\pi$ the dS temperature. Using the same color coding, we also represent stability in the energy density plot, Fig.~\ref{Edens059}(left). As may be expected, at small $H$ the globally stable state corresponds to the one with the smallest energy density.

\begin{figure}[t]
    \begin{center}
    	\includegraphics[width=0.9\textwidth]{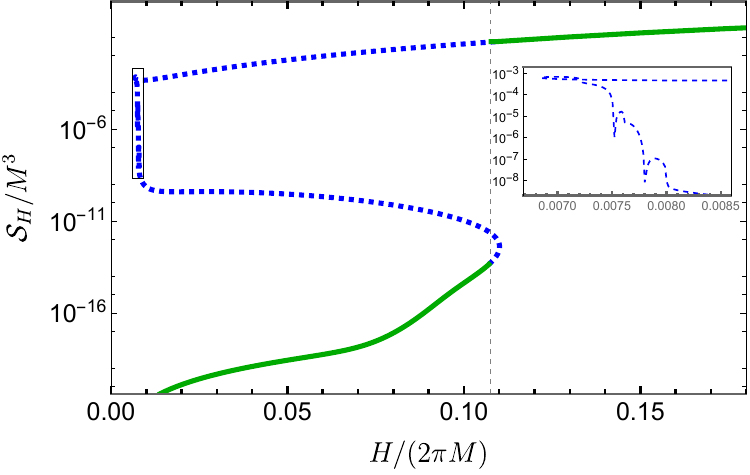}
    \end{center}
    \caption{\small\label{SAH059} 
    Entropy density per unit physical volume of the QFT in a de Sitter–invariant state for the model with $\phi_M$=0.59, extracted from the area of the apparent horizon of the dual bulk solution. 
    The branches of the phase transition are shown in solid green for globally stable states and dashed blue for globally unstable states. The dashed vertical line corresponds to the critical de Sitter temperature of the phase transition.}
\end{figure}

Although the behavior of the free energy closely parallels the thermal analysis, it is important to stress that these results do not admit a straightforward interpretation in terms of an equation of state in the thermodynamic sense. In particular, since all the solutions under consideration are dS-symmetric, they satisfy $\mathcal{E} = -\mathcal{P}$  by construction. Moreover, these dS states are intrinsically out of equilibrium, as evidenced by the growth of their comoving entropy density \eqref{coco}, which follows from a constant entropy density per unit physical volume, $\mathcal{S}_H$. This entropy is shown in Fig.~\ref{SAH059}, where we see that the different branches correspond to different rates of entropy production. 
In the figure we also observe that the lower, green branch approaches zero as $H/M \to 0$. This behavior is expected: in this limit the solution approaches the flat–space vacuum at zero temperature for any value of $\phi_M$, whose physics is governed by the CFT defined at the minimum of the scalar potential \eqref{V}. For a CFT in de Sitter space there is no entropy production \cite{Buchel:2017pto}, because dS is conformally flat and the dynamics of a CFT in dS is, up to the conformal anomaly, equivalent to that in flat space. As a consequence, the entropy density must vanish smoothly in this limit. 

The entropy shown in Fig.~\ref{SAH059} displays an interesting structure near the transition, as highlighted in the inset. We have checked that this feature is not a numerical artifact by performing convergence tests using Mathematica’s NDSolve with arbitrary-precision arithmetic. Calculations were carried out with working precisions of 30 and 40 digits, while the accuracy and precision goals were varied between 10 and 25. The relative differences between the resulting solutions are below $10^{-5}$ and decrease as the accuracy and precision goals are increased, indicating that the observed structure is numerically stable.

The fact that the entropy density shown in the figure is parametrically small in units of the characteristic scale $M$ has a related origin. As explained in~\cite{Bea:2018whf}, when $\phi_M$ approaches the critical value $\phi_M^c = 0.580822$, the potential \eqref{V} develops an inflection point between the maximum at $\phi = 0$ and the minimum at $\phi = \phi_E$, signaling the emergence of a false vacuum of the QFT in flat space.  For $\phi_M = 0.59$, the potential lies very close to this critical regime. In the presence of an exact inflection point, the physics in its vicinity would again be governed by a CFT, implying vanishing entropy production. The blue branches in the figure probe precisely the region near this would-be inflection point, and as a result the entropy production is strongly suppressed.

An important conclusion of this discussion is that, by fine-tuning $\phi_M$ toward $\phi_M^c$, we obtain models that exhibit coexisting dS-invariant states at arbitrarily small values of the ratio $H/M$, while maintaining energy differences of order $M^4$ between them. We will return to this point below, where we will show that the amount of fine-tuning required is only logarithmic in this ratio.

\subsection{Parallels with thermal states}
\begin{figure}[t]
    \begin{center}
    	\includegraphics[width=0.60\textwidth]{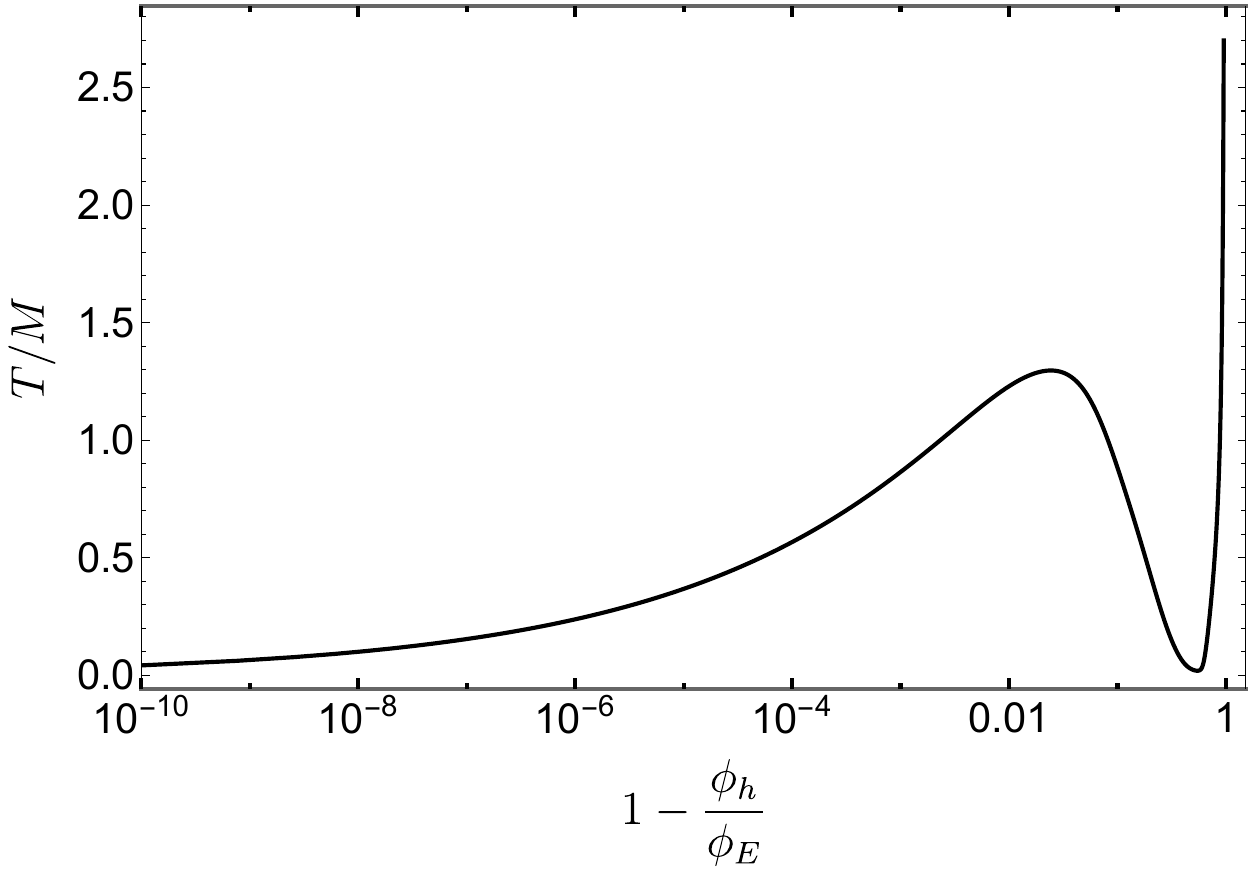}
    \end{center}
    \caption{\small\label{Tphihphiv059}
    Temperature of the field theory as a function of the scalar field at the horizon in the gravitational dual for the model with $\phi_M=0.59$. Multiple states coexist for the range of $T$ between the maximum and the minimum.}
\end{figure}

Following the discussion in Section~\ref{sec:FTgen}, from the holographic point of view the gravitational construction dual to finite-temperature states closely parallels that of dS-symmetric states: both are described by a similar metric ansatz, both possess a horizon, and the different thermal states are uniquely labeled by the value of the scalar field at the horizon, $\phi_h$. As in the dS case, the relation between the temperature and $\phi_h$ is not unique, with multiple gravitational solutions existing for the same $T$, signaling a first-order thermal phase transition. Fig.~\ref{Tphihphiv059} illustrates this structure for the model under consideration. This parallel naturally raises the question of whether the simultaneous presence of quantum and thermal phase transitions is a general feature of holographic models or specific to particular choices of the potential, a question we will return to in the next section.

\begin{figure}[!t]
    \begin{center}
    	\includegraphics[width=0.45\textwidth]{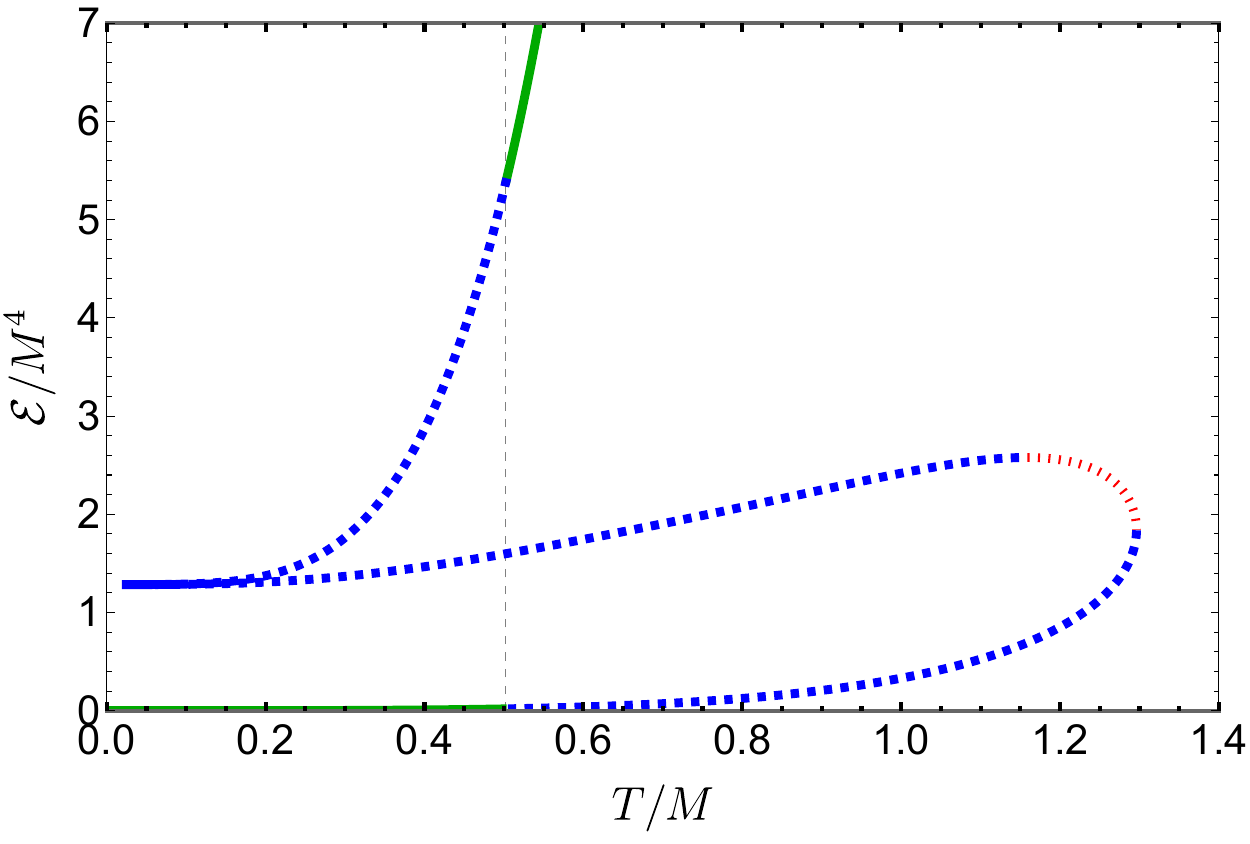}
        \includegraphics[width=0.45\textwidth]{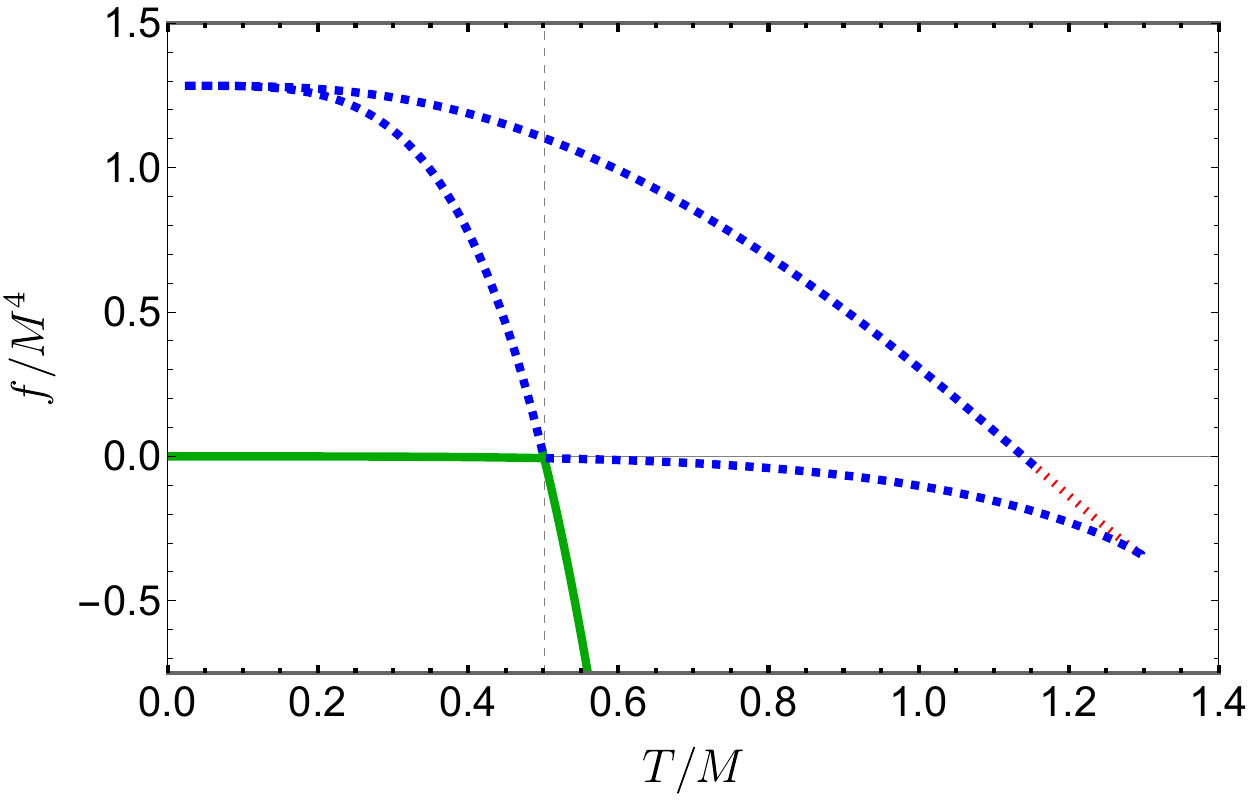}
    \end{center}
    \caption{\small\label{feT059}
    Energy density (left) and free energy density (right) as functions of the temperature for the model with $\phi_M=0.59$. The branches of the phase transition are shown in solid green for globally stable states, dashed blue for locally stable states but globally unstable, and dotted red for locally unstable states. The dashed vertical line is the critical temperature $T_c$ of the phase transition.}
\end{figure}

Fig.~\ref{feT059}(left) shows the energy density of the thermal states in this model as a function of the temperature. As in the dS case, multiple thermal states coexist down to a temperature  $T \ll \mm$, and the difference in energy density between the highest and lowest states remains of order one in units of the characteristic scale of the theory, $\mm^4$. In contrast to the dS case, the only ambiguity in the energy density is fixed by requiring that the energy of the lowest state vanishes at zero temperature. Unlike dS-symmetric states, the energy density and pressure are no longer constrained to obey $\mathcal{E}=-\mathcal{P}$. Thermodynamic relations now imply that the free energy per unit volume satisfies $f=-\mathcal{P}$. The free energy of the thermal states is shown in Fig.~\ref{feT059}(right), displaying the characteristic swallow-tail structure associated with a first-order phase transition.

As in the dS case, the globally preferred thermal state at a given temperature corresponds to the solution with the lowest free energy. However, unlike the dS-symmetric states, here we can use standard thermodynamic relations to assess the local stability of the other coexisting states.
In particular, the slope of the energy density as a function of temperature determines the specific heat: when this slope becomes negative, the specific heat turns negative, signaling a local thermodynamic instability. 
This enables a clear distinction between locally stable but globally unstable states (shown as dashed blue curves) and locally unstable states (shown as dotted red curves). Both types of states are indicated in the corresponding energy density and free energy plots.

From the free energy we can identify the critical temperature of the first-order thermal phase transition, which occurs at \(T_c/\mm \simeq 0.501\). Although we do not have direct access to the microscopic degrees of freedom underlying each phase, the holographic description reveals a clear parallel with the curvature-driven transitions. In both cases, the existence of multiple phases is tied to the structure of the vacuum RG flow of the theory: it features an extended nearly flat region near the would-be inflection point of the bulk potential,
followed by an RG trajectory controlled by the IR fixed point. In particular, for both low temperature and low curvature the lowest-energy branch of solutions is governed by the IR fixed point, whereas the high-energy branches probe the nearly flat portion of the flow without reaching the IR regime. Despite these similarities, the two transitions occur at different scales: the critical temperature of the thermal transition and the dS temperature associated with the curvature-driven transition differ by approximately a factor of five. We will return to this difference and its implications in Section~\ref{quantum-vs-thermal}.

\subsection{Quantum versus thermal phase transitions}
\label{quantum-vs-thermal}

Having established the existence of multiple dS–symmetric states in our holographic model and their interpretation as curvature-driven phase transitions, it is natural to ask how these phenomena compare to the more familiar case of thermal phase transitions. Both situations involve multiple coexisting gravitational solutions and first-order transitions, yet the underlying physics and their interpretation in the dual field theory are distinct. In this section, we highlight these parallels and differences, setting the stage for the cosmological implications that follow. To make the comparison concrete, we begin by examining how the phase transitions emerge in both cases as a function of the model parameters.

\begin{figure}[t]
    \begin{center}
    	\includegraphics[width=0.47\textwidth]{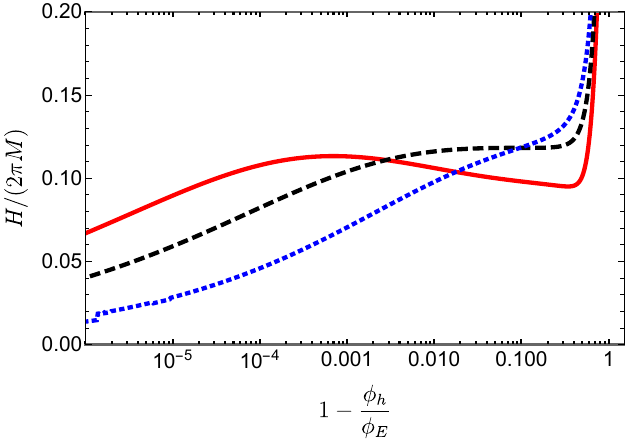}
        \hspace{5mm}
        \includegraphics[width=0.46\textwidth]{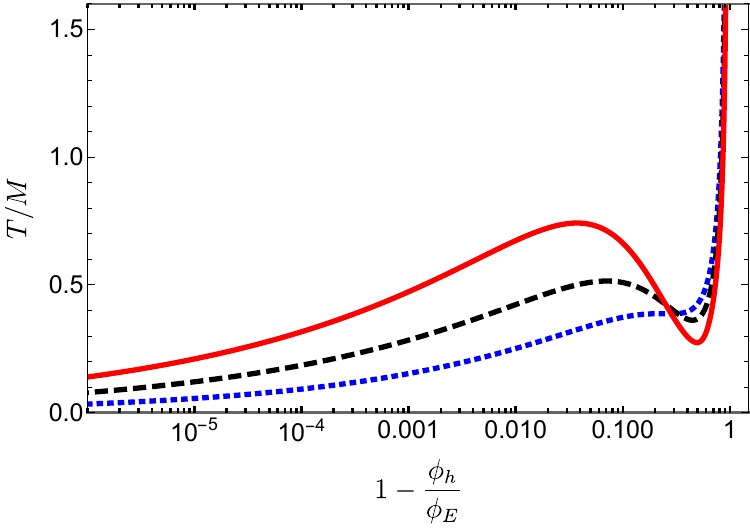}
    \end{center}
    \caption{\small\label{TvsphiH} Hubble rate (left) and temperature (right) of the field theory as a function of the scalar field at the horizon for the models with  $\phi_M=1.08$ (dotted blue),  $\phi_M=0.827$ (dashed black) and $\phi_M=0.7$ (solid red). The maximum value of $\phi_h$ (corresponding to the minimum of the potential, $\phi_E$) is different for each model.}
\end{figure}

In Fig.~\ref{TvsphiH}(left), we show the relationship between the value of the scalar field at the horizon, \(\phi_h\), and the associated expansion rate in the dual theory, \(H/\mm\), for three different choices of the bulk potential. 
For convenience, the vertical axis is expressed in terms of the dS temperature, \(T_\textrm{dS} = H / 2\pi\), which in this context serves as the parameter controlling the phase structure.

For the model with \(\phi_M=0.7\), the relation between \(\phi_h\) and \(T_\textrm{dS}\) exhibits both a maximum and a minimum, implying the presence of a quantum phase transition driven by spacetime curvature. For \(\phi_M < 0.7\), this behavior persists, with the extrema becoming more pronounced---see Fig.~\ref{Hphihphiv059}. In contrast, as \(\phi_M\) increases, the extrema move closer together until they merge at \(\phi_M = 0.827\). Beyond this value, the relation between the scalar field at the horizon and the dual field theory curvature becomes monotonic: in this regime, each value of \(H/\mm\) corresponds to a unique value of \(\phi_h\). We therefore conclude that, for this class of models  a curvature-driven quantum phase transition exists only for \(\phi_M < 0.827\).

A similar analysis can be performed for the gravitational solutions dual to thermal states of the strongly coupled field theory. While the qualitative features of the \(T-\phi_h\) relation resemble those in the dS case, there are important quantitative differences. In Fig.~\ref{TvsphiH}(right), we show the relation between \(T/\mm\) and \(\phi_h\) for the thermal states for the same holographic models. In this case, both models with \(\phi_M=0.7\) and \(\phi_M=0.827\) exhibit two extrema, indicating that the corresponding strongly coupled field theory undergoes a thermal phase transition. 
As shown in the figure, these extrema merge at \(\phi_M=1.08\). Therefore, all models with \(\phi_M < 1.08\) exhibit a thermal phase transition.

From comparing the two curves, we can draw two main conclusions. First, the existence of a thermal phase transition in the strongly coupled field theory does not necessarily imply the presence of a curvature-driven phase transition. Specifically, all models satisfying
\[
0.827 < \phi_M < 1.08
\]
serve as examples of theories with a thermal phase transition but no quantum (curvature-induced) phase transition.

Second, for models with \(\phi_M < 0.827\), where both quantum and thermal phase transitions are present, an important distinction arises. For example, for \(\phi_M=0.7\), the range of temperatures where multiple homogeneous thermal states exist is \(0.274 \leq T/\mm \leq 0.742\), while the range of de Sitter temperatures admitting multiple de Sitter-symmetric solutions is \(0.095 \leq T_\textrm{dS}/\mm \leq 0.113\). This distinction shows that, although \(T_\textrm{dS}\) may play the role of a temperature for some observables, it does not reliably predict the phase structure of dS-symmetric states.

\begin{figure}[t]
    \begin{center}
    	\includegraphics[width=0.46\textwidth]{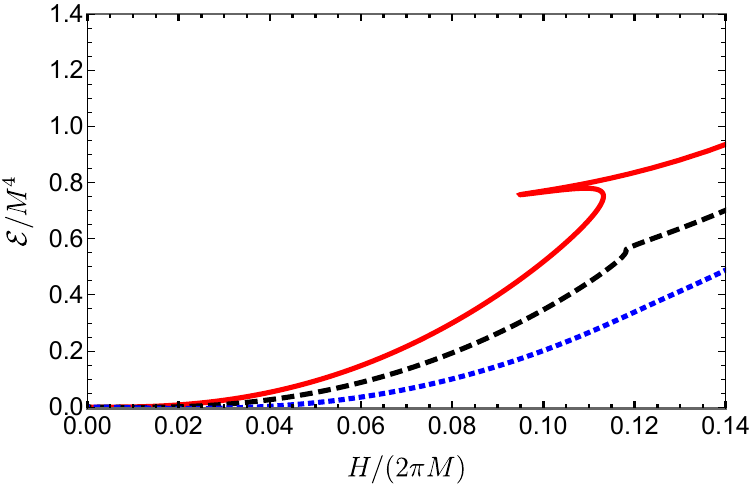}
        \hspace{5mm}
        \includegraphics[width=0.47\textwidth]{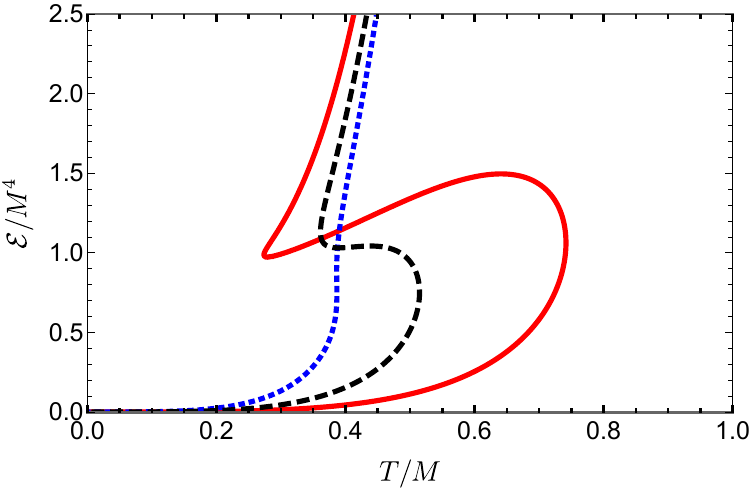}
    \end{center}
    \caption{\small\label{EoLvsH} Energy density of dS-invariant states as a function of the Hubble rate (left) and of the thermal states as a function of the temperature (right) for the models with \(\phi_M = 0.7\) (solid red), \(\phi_M = 0.827\) (dashed black), and \(\phi_M = 1.08\) (dotted blue).\label{fig:energy-density-comparison}}
\end{figure}

In Fig.~\ref{fig:energy-density-comparison}, we show the energy density of the dS-symmetric states as a function of the expansion rate \(H\) (left), and the energy density of the thermal states as a function of the temperature \(T\) (right), for the three values of \(\phi_M\) discussed above. These plots confirm that, whenever a phase transition occurs, the corresponding energy density becomes multivalued, reflecting the coexistence of distinct states. They also show that the curvature-driven and thermal transitions take place at comparable values of the energy density.  

A notable feature in the curvature-driven case, present in all models including those discussed in Section~\ref{curvature-driven-phases} (see Fig.~\ref{Edens059}), is that at small \(H\) the energy density initially decreases as the curvature increases, potentially becoming negative, before returning to positive values. This behavior is scheme-dependent: the absolute shape of the energy density curves can be modified by the choice of the renormalization parameter \(\alpha\). Nevertheless, as emphasized earlier, while the absolute values are ambiguous, the energy density differences between coexisting states at the same \(H\) or \(T\) remain unambiguous.

While the relations analyzed above are sufficient to determine the phase structure of the field theory, additional insight can be gained by examining geometric features of the dual gravitational solutions. A particularly useful quantity is the area of the event horizon. In the thermal  case, this area is directly related to the entropy density of the dual field theory. In the dS case, the entropy is instead associated with the area of the apparent horizon; nevertheless, the event-horizon area continues to encode valuable information about possible curvature-driven, quantum phase transitions in the QFT.

\begin{figure}[t]
    \begin{center}
    \includegraphics[width=0.46\textwidth]{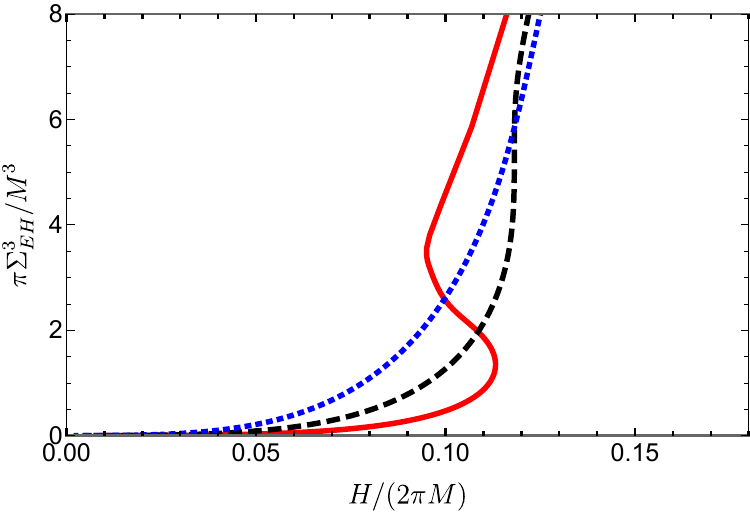}
    \hspace{5mm}
    \includegraphics[width=0.47\textwidth]{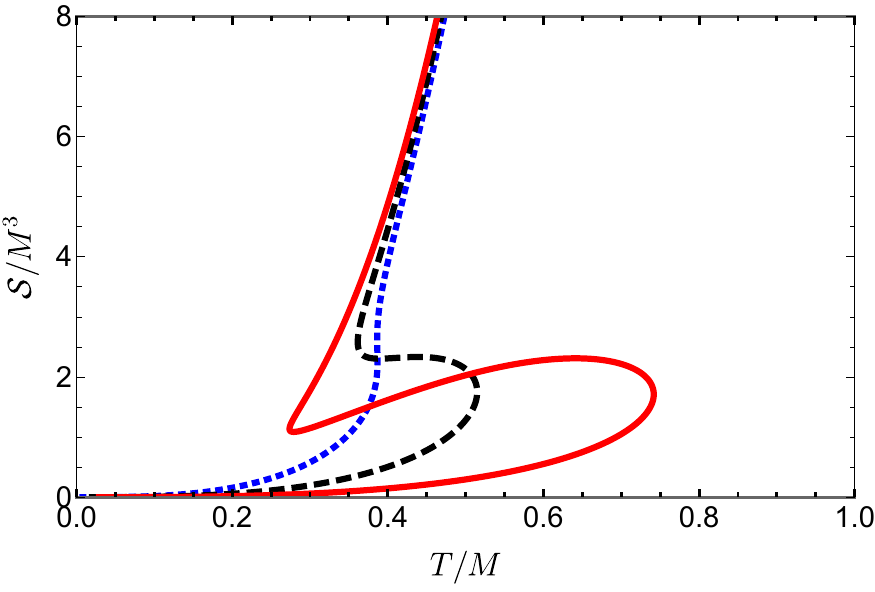}
    \end{center}
    \caption{\small\label{EHarea} Area density of the event horizon for the models with  \(\phi_M = 0.7\) (solid red), \(\phi_M = 0.827\) (dashed black), and \(\phi_M = 1.08\) (dotted blue) for the dS case (left) and for the thermal case (right).}
\end{figure}

In Fig.~\ref{EHarea}, we display the area density of the event horizon for the three models discussed previously. In the thermal case (right), this area determines the entropy density---see Eq.~\eqref{ss}. In models exhibiting phase transitions, this quantity becomes multivalued. The values of  
\(\phi_M\) at which these multivalued regions appear match the locations of the phase transitions identified in the $H - \phi_h$ and $T - \phi_h$ diagrams. Moreover, when both types of transitions occur in the same model, the ranges of \(H\) and \(T\) corresponding to multiple states are comparable, as are the magnitudes of the event horizon areas. 
\begin{figure}[!!t]
    \begin{center}
    	\includegraphics[width=0.45\textwidth]{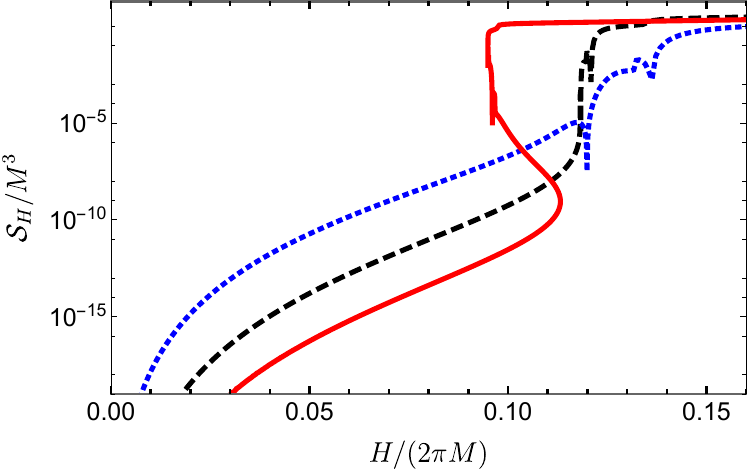}
        \hspace{5mm}\includegraphics[width=0.46\textwidth]{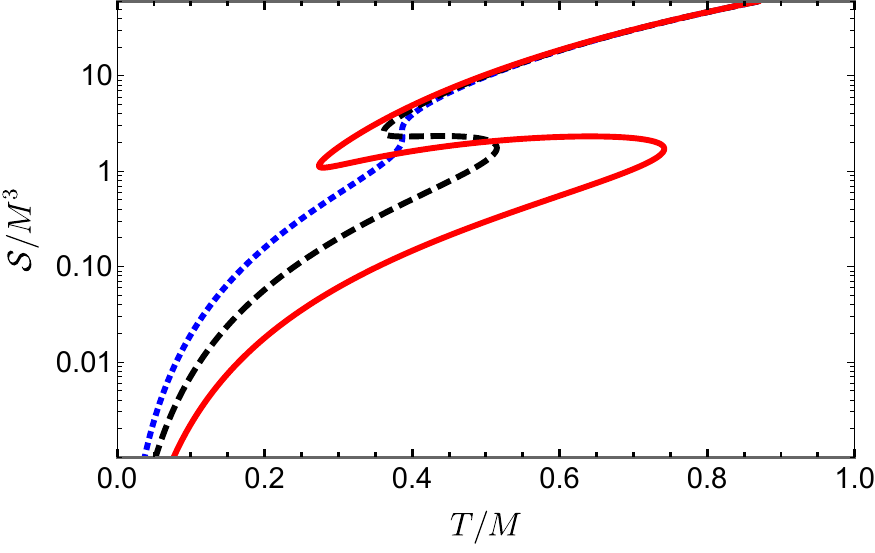}
    \end{center}
    \caption{\small\label{AHarea} Left: Entropy density of 
    dS-invariant states in the models with  \(\phi_M = 0.7\) (solid red), \(\phi_M = 0.827\) (dashed black), and \(\phi_M = 1.08\) (dotted blue).
    Near the transition, the curves exhibit interesting structure similar to that in Fig.~\ref{SAH059}.
    Right: 
    Entropy density of thermal states (same as in Fig.~\ref{EHarea}), shown in logarithmic scale for direct comparison.
    }
\end{figure}

To extract the entropy density in the dS case, we evaluate the area of the apparent horizon. Fig.~\ref{AHarea} (left) shows the 
apparent-horizon area density, \(\Sigma(u_{\rm AH})^3\), for the same three models. As with other observables, this quantity becomes multivalued whenever multiple dS-symmetric solutions coexist. Its magnitude is also significantly smaller than the corresponding event-horizon area density---both in the dS case and in the thermal case, where the relevant area is that of the event horizon. For comparison, Fig.~\ref{AHarea}(right) displays the apparent-horizon area density for thermal states. In this case, the apparent and event horizons coincide, so the curve is identical to that in Fig.~\ref{EHarea}(right), but plotted on a logarithmic scale to facilitate comparison with Fig.~\ref{AHarea}(left).

In the dS case, both horizon areas vanish in the \(H \to 0\) limit, consistent with the IR conformal fixed point of the holographic model. Across the quantum phase transition, the apparent-horizon area for dS states varies by more than ten orders of magnitude, whereas the thermal entropy changes by only about two orders of magnitude. This contrast reflects the fact that these quantities probe different---though related---features of the system. In a given thermal state, the entropy is constant, whereas in dS-invariant states the apparent-horizon area monitors entropy production. Its sharp jump across the transition therefore provides the most striking signature of the qualitatively different dynamics characterizing the coexisting dS–symmetric branches.

\section{Cosmological dynamics 
}
\label{sec:cosmo}
In this section we  examine what the universe would look like if its matter content were that of one  of the QFTs  analyzed in this paper. In this framework, we imagine promoting gravity to a dynamical field and coupling it to the strongly interacting QFT, allowing both to evolve self-consistently. In general, this leads to a time-dependent expansion rate. Here we would like to see if a late-time, self-consistent, dS-invariant solution exists. Such a state represents a consistent cosmological solution only if its expansion rate, $H=\hcos$, satisfies  the Friedmann equation  
\begin{equation}
\label{cos}
     \hcos^2\,\,\,=\,\,\,
     \frac{1}{3\mpl^2}\, \rho(\hcos) \,\,\,=
     \,\,\,\frac{1}{6\pi^2\mef^2}\, \mathcal{E}(\hcos)
\end{equation}
with the energy density on the right-hand side given by the holographic analysis at fixed expansion rate. As discussed in Section \eqref{scheme}, in the semiclassical approach the ambiguities in the definition of the stress tensor can be absorbed into redefinitions of the cosmological constant and the species scale~\cite{Birrell:1982ix}. Different choices of renormalization scheme in \eqref{cos} then lead to the same physical expansion rate. For concreteness, we  perform our analysis in the scheme defined by \eqref{special} and \eqref{aa}. In this scheme, our physical assumption that flat space is a solution of the QFT coupled to dynamical gravity is encoded in the absence of a cosmological constant term in \eqref{cos}.

\begin{figure}[!!t]
    \begin{center}
    	\includegraphics[width=0.45\textwidth]{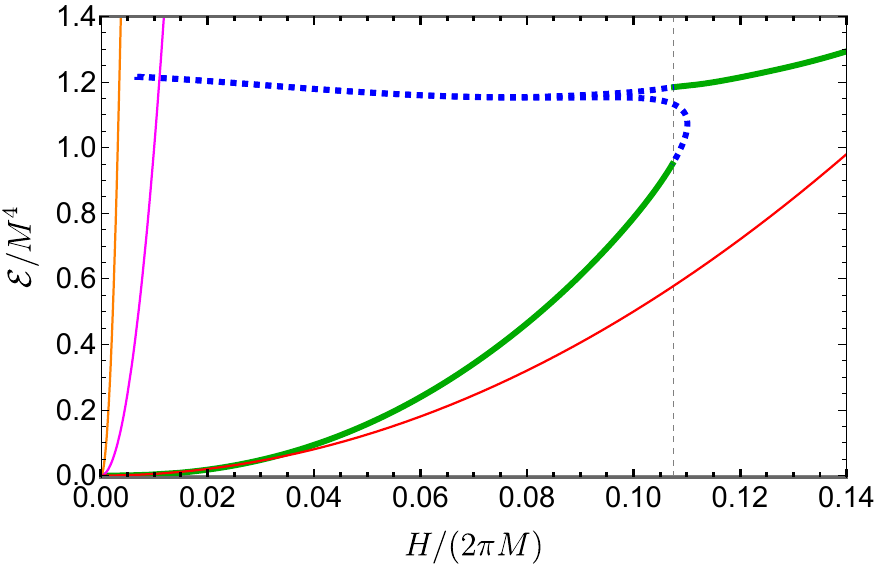}
        \hspace{5mm}\includegraphics[width=0.46\textwidth]{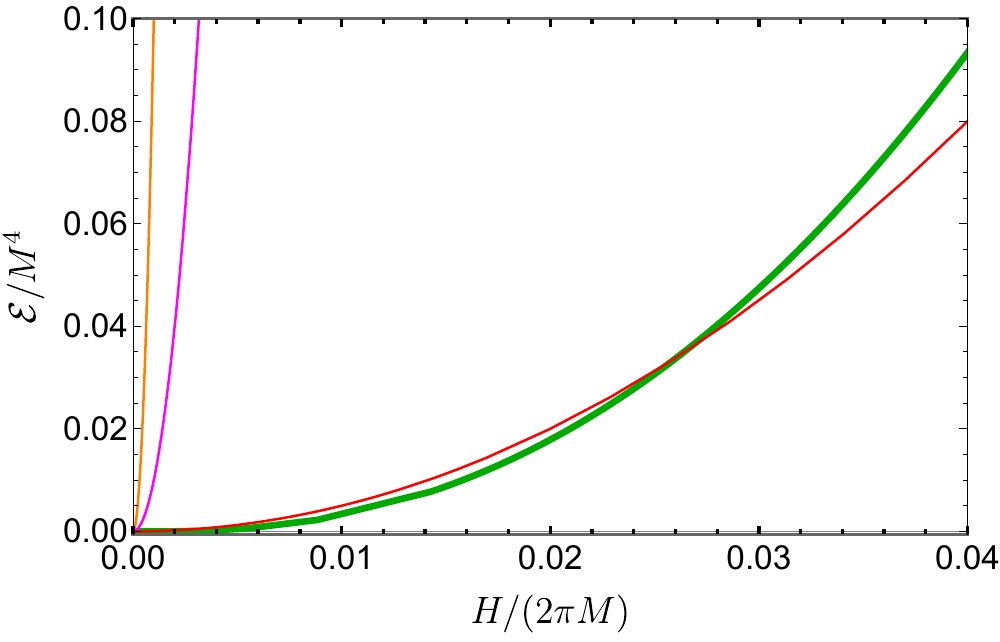}
    \end{center}
    \caption{\small Energy density of dS-invariant states of the QFT corresponding to \mbox{$\phi_M=0.59$} (the same as in Figs.~\ref{Edens059_D} and \ref{Edens059}), together with three parabolas with coefficients 
    $10^5$ (orange), $10^4$ (magenta) and 50 (red).}
    \label{pepe}
\end{figure}

The holographic analysis of the previous sections has already identified all dS–symmetric states available at a given expansion rate. The remaining task is therefore to determine which of these states are compatible with the Friedmann equation.
These solutions can be visualized as the intersection points between the parabola in the $(H,\mathcal{E})$-plane defined by Eq.~\eqref{cos} and the curve $\mathcal{E}=\mathcal{E}(H)$, which corresponds to dS--invariant states of the QFT with expansion rate $H$. In Fig.~\ref{pepe}, we show this curve for the model with $\phi_M=0.59$, together with three parabolas corresponding to different values of the ratio $\msp/M$.

Consider first solutions that lie on the lower green branch, as shown Fig.~\ref{pepe}(right). In our renormalization scheme, the energy density on this branch scales as 
\beq
\mathcal{E} \sim O(H^2 M^2) \,.
\label{green}
\eeq
Substituting this behaviour in \eqref{cos} immediately leads to the conclusion that, for a solution to exist on this branch, we must have 
\beq
\hcos \sim M \sim \msp \,, 
\label{out}
\eeq
which lies outside the semiclassical regime \eqref{semi}.  The situation in this case is analogous to that in the trace-anomaly driven scenario \cite{Starobinsky:1980te}.

\begin{figure}[!!t]
    \begin{center}
            	\includegraphics[width=0.75\textwidth]{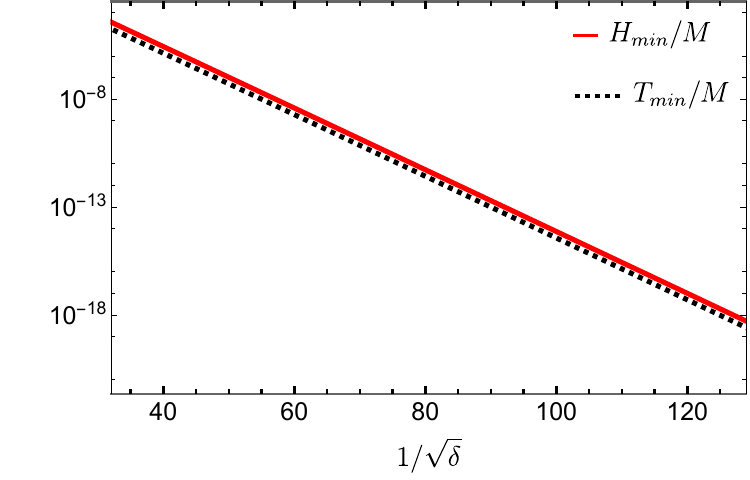}
    \end{center}
    \caption{\small\label{fig:HTminplot} Behavior of our family of holographic models as $\delta = \phi_M - \phi_M^c \to 0^+$. The solid red line indicates the minimum Hubble rate, $H_{\min}$, at which multiple de Sitter--symmetric states exist, while the dashed black line shows the minimum temperature, $T_{\min}$, down to which multiple thermal states persist.
}
\end{figure}

Consider now the possible solutions on the blue branches. Along these branches, the energy density scales as $\mathcal{E}\sim M^4$ at small $H$. Substituting this behaviour in \eqref{cos}, we obtain solutions satisfying 
\begin{equation}
\frac{H_{\rm cosmo}}{\mm} \sim \frac{\mm}{\mef} \,.
\end{equation}
For these solutions to lie within the semiclassical regime, both ratios must be parametrically small. Since the lowest accessible value of $H$ on the blue branches is $\hmin$, this condition requires  the ratio $\hmin/M$ to  be sufficiently small, as illustrated by the magenta curve in Fig.~\ref{pepe}. Otherwise, no solution exists, as in the case of the orange curve. In our family of holographic models, this requirement can be satisfied with only logarithmic fine-tuning. Specifically, as $\phi_M$ approaches the critical value  $\phi_M^{c}\simeq 0.580822$ from above, $\hmin/M$ scales as 
\beq
\label{expexp}
\frac{\hmin}{M} \sim \exp 
\left( - \frac{b}{\sqrt{\delta}} \right)
\,, \qquad b \simeq 0.329
\,,
\eeq
where 
\beq
\delta \equiv \phi_M - \phi_M^c \,.
\eeq
This exponentially rapid decrease of $\hmin/M$ as $\delta \to 0^{+}$ is illustrated by the red solid  line in Fig.~\ref{fig:HTminplot}. The origin of this exponential dependence can be understood as follows. As $\delta \to 0^{+}$, the potential approaches the formation of an inflection point, as shown in the inset of Fig.~\ref{potpot}. Consequently, the bulk geometry develops a long, approximately AdS throat in the vicinity of this point~\cite{Bea:2018whf}. In this region, the relation between proper distance along the holographic direction and the corresponding energy scale in the dual QFT is exponential, a feature familiar from other contexts such as the Randall--Sundrum resolution of the hierarchy problem~\cite{Randall:1999ee}. In the present case, the length of the AdS throat scales as $1/\sqrt{\delta}$, while the QFT energy scale dual to the point where the throat terminates is given by $\hmin$, leading directly to Eq.~\eqref{expexp}. In summary, what may appear as a large hierarchy in the QFT can be achieved via a logarithmic fine-tuning of the holographic parameter, $\phi_M$. 

The presence of an inflection point at $\phi_M = \phi_M^{c}$ also has implications for the thermal structure of the QFT in flat space. As $\delta \to 0^{+}$, metastable thermal states with progressively larger amounts of overcooling appear, persisting down to a minimum temperature $T_{\min}$. This temperature marks the onset of the spinodal instability. As indicated by the dashed line in Fig.~\ref{fig:HTminplot}, the value of $T_{\min}$ exhibits a similarly sharp dependence on $\delta$ as $H_{\min}$. For all models in our family, we find
\beq
\frac{H_{\min}}{2\pi} < T_{\min} < H_{\min}\,.
\eeq
In the limiting case $\delta = 0$, a false vacuum appears in the theory.

\begin{figure}[t]
    \begin{center}
    	\includegraphics[width=0.60\textwidth]{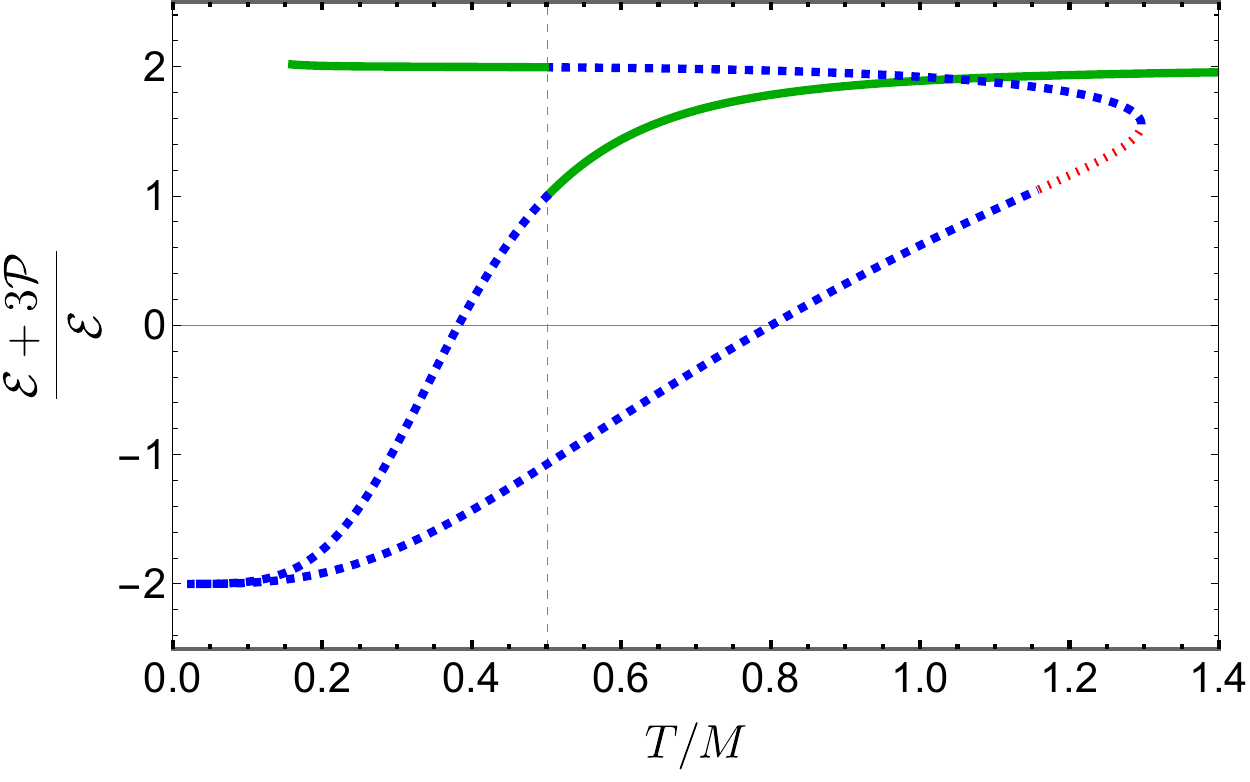}
    \end{center}
    \caption{\small\label{fig:e3pplot} Energy density plus three times the pressure, normalized by the energy density, for thermal states in flat space of the QFT corresponding  to  $\phi_M = 0.59$. 
A universe filled with this matter undergoes accelerated expansion whenever this quantity becomes negative. Different branches correspond to stable (solid green), metastable (dashed blue) and unstable (dotted red) states. 
}
\end{figure}

These overcooled states satisfy the necessary conditions for thermal inflation. This is illustrated in Fig.~\ref{fig:e3pplot}, where $\mathcal{E} + 3\mathcal{P}$ becomes negative in the temperature range where metastable thermal states exist. As the temperature decreases, the pressure approaches $\mathcal{P} \simeq -\mathcal{E}$. In this regime, the universe undergoes nearly exponential expansion, with a scale set by $H_{\rm cosmo} \sim \mm^2 / \mef$, since the energy density remains of order $\mm^4$.

We can now envision a dynamical path through which our dS-invariant states may be realized during cosmological evolution. Consider a model with $H_{\rm min} < H_{\rm cosmo}$ and an initial state with sufficiently large energy density on the green branch on the right-hand side of Fig.~\ref{fig:e3pplot}. At early times, the evolution gives rise to an FLRW universe in which matter is in local thermal equilibrium at a temperature $T > H$ and can be described using the flat-space equation of state. As the universe expands and the plasma cools, it enters the regime of thermal inflation. As seen in Fig.~\ref{fig:e3pplot}, this occurs when $T \sim M$. In this regime, the expansion becomes nearly exponential with an almost constant rate $H \simeq H_{\rm cosmo}$. Over a number of e-folds of order $\log(\mm/\hcos) \sim \log(\mef/\mm)$, the temperature decreases to $T \sim \hcos > \hmin> T_{\rm min}$. Beyond this point, the plasma relaxation rate becomes slower than the expansion rate, and the matter falls out of equilibrium \cite{Casalderrey-Solana:2020vls}. Consequently, the subsequent evolution cannot be predicted based on the equilibrium phase diagram of Fig.~\ref{fig:e3pplot}. Our analysis then identifies a natural attractor for this late-time evolution: an exact dS-invariant state in which the expansion is 
supported by the stress tensor generated by the expansion itself. We will come back to the attractor nature of this solution in the next section.

\section{Discussion}
\label{discussion}
Our main result is that, in our holographic setup, multiple dS-symmetric states  arise at arbitrarily small Hubble rates, requiring only logarithmic fine tuning of the model parameter. As explained below \eqref{coco}, the area growth of the apparent horizon of the dual solutions indicates that the entropy of these states increases with time, supporting their out-of-equilibrium nature. Presumably, in the QFT this entropy must be interpreted as a coarse-grained quantity, reflecting the progressive loss of information about the initial conditions due to cosmological expansion. This interpretation would closely parallel the coarse-grained entropy associated with apparent horizons in holographic black holes \cite{Engelhardt:2017aux}. 

Within our model, we find that the appearance of  small-curvature states is correlated with proximity in parameter space to a false vacuum in flat space. This observation indicates that the existence of these states is not peculiar to holography, but instead reflects a more general property of QFTs near the point at which a false vacuum disappears. Similarly, curvature-driven and thermal phase transitions exhibit comparable dependence on the parameters of the model. Despite these parallels in our one-parameter setup, preliminary investigations suggest that, in more general holographic models, the properties of the QFT in dS and in flat space can differ qualitatively.

Although we have focused on dS-invariant states, our analysis suggests that such states can be reached dynamically through cosmological evolution, once the initial matter content has fallen out of equilibrium. This expectation is supported by the numerical studies of \cite{Buchel:2017pto,Casalderrey-Solana:2020vls}. These references showed that,  when the QFT is initialized in a highly excited configuration in a fixed dS background, it evolves at late times toward a dS-symmetric state analogous to those described here. We are currently performing numerical simulations along the lines of Ref.~\cite{Ecker:2021cvz} to establish this behavior in the case where the QFT is coupled to dynamical gravity.

The stability or lack thereof of the dS-invariant states is an interesting open question. Consideration of the Euclidean action of the solutions suggests that some of these states may be metastable or unstable. One possible decay channel consists of the nucleation of bubbles of the preferred phase. However, this process is suppressed in the $N\to \infty$ limit, as discussed in \cite{Bea:2021zol}. Even at finite $N$, the nucleation rate may remain sufficiently small that the transition fails to complete, as happens in quasi-conformal matter~\cite{Lewicki:2021xku}. In addition to this nonperturbative channel, the dS–symmetric states could also be locally unstable. The linear stability of these solutions at fixed $H$ can be analyzed following the approach of~\cite{Mashayekhi:2025jyg}.

In summary, the cosmological evolution of the strongly coupled matter described by our holographic model, when starting from a high-energy state, may naturally give rise to an inflationary phase in which the QFT evolves far from equilibrium. This period of accelerated expansion could persist for a large number of e-folds before eventually ending through one of the decay mechanisms discussed above. It is intriguing to speculate whether the actual matter content of our Universe, or perhaps a hidden dark sector, could exhibit similar behavior. However, we emphasize again that determining whether this scenario can serve as a viable model of inflation would require, among other ingredients, a detailed analysis of cosmological perturbations, a mechanism for a graceful exit, and a reheating process. We leave these important questions for future work.

\section*{Acknowledgements}
We thank Roberto Emparan, Jaume Garriga and Elias Kiritsis for discussions. The authors used OpenAI’s ChatGPT (GPT-5, October 2025 version) to assist in refining the language and reference organization of the manuscript. All scientific interpretations, results, and conclusions are solely the responsibility of the authors. 
We acknowledge financial support from grants CEX2019-000918-M and CEX2024-001451-M funded by MICIU/AEI/10.13039/501100011033, from grants No.~PID2022-136224NB-C21 and PID2022-136224NB-C22 from the Spanish Ministry of Science, Innovation and Universities, and from grant No. 2021-SGR-872 funded by the Catalan Government. The work of J.G. received the support of the fellowship LCF/BQ/DI23/11990069 from ``la Caixa'' Foundation (ID 100010434). This research is also funded by the European Union (ERC, HoloGW, Grant Agreement No. 101141909). Views and opinions expressed are, however, those of the authors only and do not necessarily reflect those of the European Union or the European Research Council. Neither the European Union nor the granting authority can be held responsible for them.

\bibliography{refs}{}
\bibliographystyle{JHEP}

\end{document}